\documentclass[12pt,a4paper]{article}
\usepackage{amsmath,multirow,amssymb,paralist,color,graphicx,setspace,longtable,subfig,microtype,bm,todonotes}
\usepackage{booktabs}
\usepackage{xcolor}
\usepackage[notablist,nofiglist,tablesfirst]{endfloat}
\usepackage[authoryear,semicolon,longnamesfirst]{natbib}
\usepackage{slashbox}
\usepackage[hmargin=1in,vmargin=1in]{geometry}
\usepackage[pdftex,colorlinks=true]{hyperref}
\hypersetup{citecolor=darkblue,linkcolor=darkblue,urlcolor=darkblue}
\definecolor{darkblue}{rgb}{0,0,.6}

\newcommand\possessivecite[1]{\citeauthor{#1}'s \citeyearpar{#1}}
\DeclareMathOperator*{\argmin}{arg\; min}

\definecolor{Rcolor}{RGB}{150,160,190}

\newcommand{\Rx}{\fontsize{10pt}{12pt}\selectfont
\raisebox{.3em}{\hspace{1.2em}%
\llap{\resizebox{1.09em}{.5em}{\color{black}$\bigcirc$}}%
\llap{\resizebox{1.199em}{.55em}{\color{darkgray}$\bigcirc$}}%
\llap{\resizebox{1.19em}{.52em}{\color{gray!50}$\bigcirc$}}%
\llap{\resizebox{1.1em}{.5em}{\color{gray}$\bigcirc$}}%
\llap{\resizebox{1.25em}{.55em}{\color{gray}$\bigcirc$}}%
}%
\hspace{-.85em}%
\textbf{%
\textcolor{black}{\textsf{R}}%
\hspace{-.025em}\raisebox{.01em}{\llap{\textcolor{Rcolor}{\textsf{R}}}}%
}}%

\newbox\rbox
\savebox\rbox{\scalebox{0.1}{\Rx}}

\begin{document}

\begin{center}
 \large Bayesian bandwidth estimation for a nonparametric functional regression model with mixed types of regressors and unknown error density
\end{center}

\begin{center}
 Han Lin Shang\footnote{Postal address: Research School of Finance, Actuarial Studies and Applied Statistics, Australian National University, Kingsley Street, Canberra, ACT 0200, Australia; Telephone: +61 2 6125 0535, Fax: +61 2 6125 0087; Email: hanlin.shang@anu.edu.au} \\
Australian National University
\end{center}

\vspace{0.5in}

\begin{abstract}

We investigate the issue of bandwidth estimation in a nonparametric functional regression model with function-valued, continuous real-valued and discrete-valued regressors under the framework of unknown error density. Extending from the recent work of \citet[][Computational Statistics \& Data Analysis]{Shang13}, we approximate the unknown error density by a kernel density estimator of residuals, where the regression function is estimated by the functional Nadaraya-Watson estimator that admits mixed types of regressors. We derive a kernel likelihood and posterior density for the bandwidth parameters under the kernel-form error density, and put forward a Bayesian bandwidth estimation approach that can simultaneously estimate the bandwidths. Simulation studies demonstrated the estimation accuracy of the regression function and error density for the proposed Bayesian approach. Illustrated by a spectroscopy data set in the food quality control, we applied the proposed Bayesian approach to select the optimal bandwidths in a nonparametric functional regression model with mixed types of regressors.

  \vspace{.2in}



\end{abstract}

\newpage

\section{Introduction}\label{sec:1}

Recent advances in computer recording and storing technology facilitate the presence of functional data sets, motivated many researchers to consider various functional regression models for estimating the relationship between predictor and response variables, where at least one variable is functional in nature. The first functional formulation of a linear model dates back to a discussion by \cite{HM93}, and it is later extended in detail by \cite{RS05}. Since then, it has been further extended or modified to take into account possible nonlinear relationship between predictor and response variables. Some of the extended regression models include the functional logit regression model \citep{AEV08},  functional polynomial regression model \citep{YM10}, functional additive regression model \citep{FG13}, and nonparametric functional regression model \citep{FV06}, semi-functional partial linear model \citep{AV06, AV08}, to name only a few. Due to the fast development in functional regression models, it has received an increasing popularity in various fields of application, such as age-specific mortality and fertility forecasting in demography \citep{HS09}, analysis of spectroscopy data in chemometrics \citep{FV02},  calculations of conditional value-at-risk and expected shortfall in econometrics \citep{FQ14}, earthquake modelling \citep{DFV11}, and ozone level prediction \citep{QF11}.

Despite the fast development in functional regression models for finding the relationship between predictor and response variables, the type of predictor variable is often limited to a function-valued variable and error density is often assumed to be Gaussian. In the field of functional regression models, error density estimation remains largely unexplored. However, the estimation of error density is important to understand the residual behaviour and to assess the adequacy of error distribution assumption \citep[see for example,][]{AV01,CS08}; the estimation of error density is useful to test the symmetry of the residual distribution \citep[see for example,][]{AL97,ND07}; the estimation of error density is important to statistical inference, prediction and model validation \citep[see for example,][]{Efromovich05, MN10}; and the estimation of error density is useful for the estimation of the density of the response variable \citep[see for example,][]{EJ12}. In the realm of financial asset return, an important use of the estimated error density is to estimate value-at-risk for holding an asset. In such a model, any wrong specification of the error density may produce an inaccurate estimate of value-at-risk and make the asset holder unable to control risk. Therefore, being able to estimate the error density is as important as being able to estimate the regression function.

This motivates the investigation of a kernel-form error density for estimating unknown error density in a nonparametric functional regression model, where the response variable is scalar but the predictor variables can be function-valued, continuous real-valued and discrete-valued. The contribution of this paper is two fold: First, to deal with the mixed types of regressors, we consider three types of kernel functions to model the explanatory power of the function-valued, continuous real-valued and discrete-valued predictors on the scalar real-valued response. Second, to accurately estimate the error density, we put forward the kernel-form estimator which depends on three parameters: 1) the type of semi-metric used to measure distances among curves, such as a semi-metric based on second-order derivative or a semi-metric based on functional principal components; 2) residuals fitted through the functional Nadaraya-Watson (NW) estimator of the regression function; 3) and the bandwidth of residuals. \cite{Cheng02, Cheng04} studied weak and strong uniform consistency of such an error density estimator, while \cite{Samb11} established the optimal convergence rate of the kernel-form error density estimator in a multivariate framework. Recently, \cite{Shang13} proposed a novel Bayesian bandwidth estimation procedure to simultaneously estimate the bandwidths in the functional NW estimator of the regression function and the kernel-form error density in a nonparametric functional regression function with only the functional predictor. In this paper, we extend this Bayesian approach to the same nonparametric functional regression model with mixed types of regressors.

\section{Bayesian bandwidth estimation}\label{sec:2}

We present the basics of this method and refer readers to the textbooks by \cite{FV06} on nonparametric functional regression and \cite{LR07} on nonparametric multivariate regression with mixed types of regressors, respectively. The essential idea of functional NW kernel smoothing is to allow flexible functional estimation of the unknown regression function. For a scalar real-valued response $y_i$, we consider the following nonparametric functional regression model,
\begin{equation}
  y_i = m(\bm{z}_i)+\varepsilon_i,\qquad i=1,2,\dots,n,\label{eq:1}
\end{equation}
where $m$ is a smooth function from a square-integrable function space to the real line, $\bm{z}_i = (T_i, \bm{X}_i^{\text{(c)}}, \bm{X}_i^{\text{(d)}})$ with $T_i \in L[0,1]$ being an infinite-dimensional functional variable bounded within a compact interval, $\bm{X}_i^{\text{(c)}}$ being $p$-dimensional continuous real-valued variables and $\bm{X}_i^{\text{(d)}}$ being $q$-dimensional discrete-valued variables, $(\varepsilon_1,\varepsilon_2,\dots,\varepsilon_n)$ are independent and identically distributed (iid) errors with unknown density denoted as $f(\varepsilon)$. The discrete-valued predictors can be either ordered or unordered, which in turn affects the selection of suitable kernels (see equations~\eqref{eq:AA76} and~\eqref{eq:LR07}). The regression function is denoted by $m(\bm{z}_i)=\text{E}(y|\bm{z})$, and we assume that $E(\bm{z}|\varepsilon)=0$. We investigate the problem of nonparametric estimation of the probability density function of the error term, which in turn might provide a good estimate of regression function. As noted by \cite{Samb11}, the difficulty of estimating error density is the fact that the regression error term is not observed and thus must be estimated.

The flexibility of model~\eqref{eq:1} stems from the fact that the unknown regression function $m(\cdot)$ does not need to have a specific functional form. With some smoothness properties, the functional form of $m(\cdot)$ is often estimated in a data-driven manner. There are a growing amount of literature on the development of nonparametric functional estimators, such as functional NW estimator \citep{FV06}, functional local linear estimator \citep{BEM11}, functional $k$-nearest neighbour estimator \citep{BFV09} and distance-based local linear estimator \citep{BDF10}. Throughout the paper, we demonstrate the proposed method using the functional NW estimator because of its simplicity and mathematical elegance.

\subsection{Functional Nadaraya-Watson (NW) estimator}

The functional NW estimator of the conditional mean can be defined as
\begin{equation*}
   \widehat{m}(\bm{z};\delta, \bm{h},\bm{\lambda}) = \frac{\frac{1}{n}\sum^n_{i=1}W_{\delta,\bm{h},\bm{\lambda},\bm{z}_i}(\bm{z})y_i}{\frac{1}{n}\sum^n_{i=1}W_{\delta,\bm{h},\bm{\lambda},\bm{z}_i}(\bm{z})},
\end{equation*}
where $W_{\delta,\bm{h},\bm{\lambda},\bm{z}_i}(\bm{z})=K_{\delta}(d(T_i,T))\times K_{\bm{h}}^{\text{(c)}}(\bm{X}_i^{\text{(c)}}-X^{\text{(c)}})\times K_{\bm{\lambda}}^{\text{(d)}}(\bm{X}_i^{\text{(d)}},X^{\text{(d)}})$ is a generalised product kernel that admits function-valued, continuous real-valued and discrete-valued predictors.

The kernel function for a function-valued predictor can be expressed as
\begin{equation*}
  K_{\delta}(d(T_i,T))=\frac{1}{\delta}\phi\left(\frac{d(T_i,T)}{\delta}\right),
\end{equation*}
where $K$ is a symmetric real valued function defined on $R^+$, $d(\cdot,\cdot)$ is a semi-metric used to measure distances among curves, $\delta\in R^+$ represents a bandwidth associated with an infinite-dimensional function-valued predictor, and $\phi(\cdot)$ is the second-order Gaussian kernel function. For simplicity, we consider just one function-valued predictor in this paper, although the methodology can be extended to multiple function-valued predictors \citep[see for example,][]{GBC+11}.

In the nonparametric functional regression model, faster rates of convergence can be obtained (and therefore nicer theoretical and practical results), as long as one uses a semi-metric $d$ that increases the concentration of the explanatory variable $T$, while reflecting as much as possible the effect of $T$ on the response variable \citep{DFV11}. For a non-smoothed functional data, a semi-metric based on functional principal component analysis should be considered; for a smoothed functional data, a semi-metric based on derivative should be considered. Because the data we considered are smooth, a semi-metrics based on the $2^{\text{nd}}$-order derivative of $T$ is used to measure the distance between two curves, that is given by
\begin{equation}
  d_{2}^{\text{deriv}}(T_i,T)=\sqrt{\int_t\big[T_i^{''}(t)-T^{''}(t)\big]^2dt},\label{eq:2}
\end{equation}
where $T^{''}$ being the $2^{\text{nd}}$-order derivative of $T$ \citep[see for example][]{Goutis98, FV09}. We refer readers to the work of \citet[][pp.32-33]{FV06} for the calculation of~\eqref{eq:2} in detail.

The product kernel for the continuous real-valued predictors can be expressed as
\begin{equation*}
  K_{\bm{h}}^{\text{(c)}}\left(\bm{X}_i^{\text{(c)}}-X^{\text{(c)}}\right)=\prod^p_{j=1}\frac{1}{h_j}\phi\left(\frac{X_{ij}^{\text{(c)}}-X_j^{\text{(c)}}}{h_j}\right),
\end{equation*}
where $\bm{h}=(h_1,h_2,\dots,h_p)\in R^+$ is a row vector of bandwidths associated with $p$-dimensional continuous real-valued predictors, $X_{ij}^{\text{(c)}}$ is the $j^{\text{th}}$ predictor of $\bm{X}_i^{\text{(c)}}$, $X_j^{\text{(c)}}$ is the $j^{\text{th}}$ grid point of $X^{\text{(c)}}$, and $\phi(\cdot)$ is the second-order Gaussian kernel function.

For binary variables, the choice of discrete kernel we considered is \citeauthor{AA76}'s \citeyearpar{AA76} kernel, given by
\begin{equation}
K_{\lambda_s}^{\text{(d)}}\left(X_{is}^{\text{(d)}}, X_s^{\text{(d)}}\right) = \left\{ \begin{array}{ll}
         1-\lambda_s & \mbox{if $X_{is}^{\text{(d)}} = X_s^{\text{(d)}}$}\\
         \lambda_s & \mbox{otherwise}\end{array},\qquad s=1,2,\dots,q, \right.\label{eq:AA76}
\end{equation}
where $\bm{\lambda}=(\lambda_1,\lambda_2,\dots,\lambda_q)\in [0,0.5]$ represents a row vector of bandwidths associated with $q$-dimensional discrete-valued predictors, $X_{is}^{\text{(d)}}$ is the $s^{\text{th}}$ predictor of $\bm{X}_i^{\text{(d)}}$ and $X_s^{\text{(d)}}$ represents possible discrete outcomes. Notice that when $\lambda_s=0$, the above kernel function $K_{\lambda_s}^{\text{(d)}}\left(X_{is}^{\text{(d)}}, X_s^{\text{(d)}}\right)$ becomes an indicator function which takes values 0 and 1.

So far, we consider discrete variables that do not have a natural ordering, an example of which includes gender. However, there are some cases where a discrete variable has a natural ordering, an example of which includes credit rating. For categorical variables, the choice of discrete kernel we considered is \citeauthor{LR07}'s \citeyearpar{LR07} kernel, which incorporates the natural ordering. It is expressed as
\begin{equation}
  K_{\lambda_s}^{\text{(d)}}\left(X_{is}^{\text{(d)}}, X_s^{\text{(d)}}\right) = \left\{ \begin{array}{ll}
        1 & \mbox{if $X_{is}^{\text{(d)}}=X_s^{\text{(d)}}$}\\
        \lambda_s^{\left|X_{is}^{\text{(d)}}-X_s^{\text{(d)}}\right|} & \mbox{otherwise}\end{array}, \qquad s=1,2,\dots,q, \right.\label{eq:LR07}
\end{equation}
where $\bm{\lambda}=(\lambda_1,\lambda_2,\dots,\lambda_q)\in [0,1]$ represents a row vector of bandwidths associated with $q$-dimensional discrete-valued predictors.

Having determined the types of kernel function, the unknown parameters in the functional NW estimator are the bandwidths (also known as smoothing parameters in the field of nonparametric smoothing). As it is always the case in nonparametric estimation, the role of smoothing parameters is prominent. For example, the rates of convergence of the nonparametric functional estimator can be divided into two parts: a squared bias component which increases with the bandwidths, and a variance component which decreases with the bandwidths. Therefore, there is a need to select optimal bandwidths in order to balance the trade-off between squared bias and variance \citep[see][for a recent review on the bandwidth selection in the multivariate density estimation and nonparametric regression]{HSS13}. 

In the field of nonparametric functional regression, the bandwidth associated with a function-valued regressor is commonly estimated by the so-called functional cross validation \citep[see for example,][]{FV02, FV06, RV07, BFV09, AV11}. The optimal bandwidths of the kernel estimators are obtained by minimising
\begin{equation*}
  \argmin_{(\delta,\bm{h},\bm{\lambda})} \text{CV}(\delta,\bm{h},\bm{\lambda}),
\end{equation*}
where
\begin{equation*}
  \text{CV}(\delta,\bm{h},\bm{\lambda}) = \sum^n_{i=1}\left[y_i-\widehat{m}_{-i}(\bm{z};\delta,\bm{h},\bm{\lambda})\right]
\end{equation*}
and $\widehat{m}_{-i}(\bm{z};\delta,\bm{h},\bm{\lambda})$ is the leave-one-out kernel estimator. Functional cross validation is designed to assess the predictive performance of a model by an average of certain measures for the ability of predicting a subset of curves by a model fit, after deleting just these curves from the functional data set. Under the criteria of averaged squared error, integrated squared error and mean integrated squared error, \cite{RV07} presented asymptotic optimality of the functional cross validation. In addition, functional cross validation has the appealing feature that no estimation of the error variance is required. However, since residuals affect the estimation accuracy of regression function, functional cross validation may select sub-optimal bandwidths, in turn might lead to inferior estimation accuracy of regression function for a small sample. In addition, as the number of parameters increases, functional cross validation may suffer from numerical instability. Instead, we present a Bayesian bandwidth estimation approach that simultaneously estimate the bandwidths in the regression function and error density, under the framework of mixed types of regressors and unknown error density.

\subsection{Estimation of error density}

The goal is to recover an unknown error density $f({\varepsilon})$ (on the real line) from a sample $\varepsilon_1,\dots,\varepsilon_n$ of $n$ independent random observations. Because errors are often unknown, they can be approximated by residuals that are obtained from the functional NW estimator. To avoid the selection of zero value for the bandwidth, $f(\varepsilon)$ can be approximated by the leave-one-out kernel density estimator expressed as
\begin{equation}
f(\varepsilon)\approx \frac{1}{n-1}\sum^n_{\substack{j=1\\ j\neq i}}\frac{1}{b}\phi\left(\frac{\widehat{\varepsilon}_i-\widehat{\varepsilon}_j}{b}\right),\label{eq:4}
\end{equation}
where $\widehat{\varepsilon}_i$ represents the $i^{\text{th}}$ residual, and $b$ represents the residual bandwidth.

Given a set of parameters $(\delta,\bm{h},\bm{\lambda},b)$, the kernel likelihood of $\bm{y}=(y_1,y_2,\dots,y_n)^{\top}$ can be approximated by
\begin{equation}
\widehat{L}(\bm{y}|\delta,\bm{h},\bm{\lambda},b) = \prod^n_{i=1}\widehat{f}_{-i}\left[y_i-\widehat{m}_{-i}(\bm{z};\delta,\bm{h},\bm{\lambda})\right], \label{eq:5}
\end{equation}
where $\widehat{m}_{-i}(\cdot)$ and $\widehat{f}_{-i}(\cdot)$ are the leave-one-out nonparametric estimators of the regression function and error density function of the computed residuals, respectively.

\begin{figure}[!ht]
  \centering
  \includegraphics[width=\textwidth]{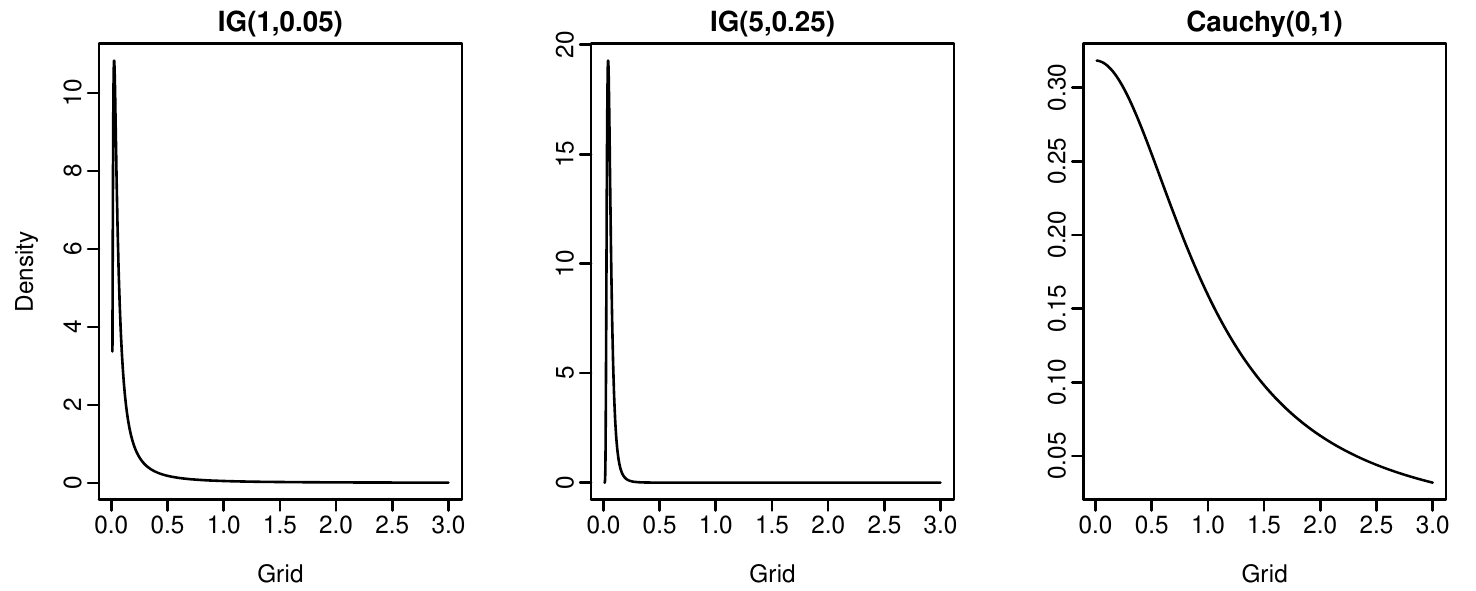}
  \caption{Probability density functions of three possible hyperparameter choices for the squared bandwidths. Throughout the paper, we use the IG(1,0.05), and report the sensitivity analysis in Table~\ref{tab:mcmctable}.}\label{fig:prior}
\end{figure}

We now discuss the issue of prior density for the bandwidths. Let $\pi(\delta^2)$, $\pi(\bm{h}^2)$, $\pi(\bm{\lambda}^2)$ and $\pi(b^2)$ be the prior of squared bandwidths $\delta$, $\bm{h}$, $\bm{\lambda}$ and $b$. Since $\delta^2$, $\bm{h}^2$ and $b^2$ play the same role as a variance parameter in the Gaussian density, we assume that the priors of $\delta^2$, $\bm{h}^2$ and $b^2$  are inverse Gamma density, denoted as IG$(\alpha_{\delta}, \beta_{\delta})$, IG$(\alpha_{h},\beta_h)$, and IG$(\alpha_b, \beta_b)$, respectively. Therefore, the prior densities of $\delta^2$, $\bm{h}^2$, and $b^2$ are
\begin{align*}
\pi(\delta^2) &= \frac{(\beta_{\delta})^{\alpha_{\delta}}}{\Gamma(\alpha_{\delta})}\left(\frac{1}{\delta^2}\right)^{\alpha_{\delta}+1}\exp\left(-\frac{\beta_{\delta}}{\delta^2}\right),\\
\pi(h_j^2) &= \frac{(\beta_{h})^{\alpha_{h}}}{\Gamma(\alpha_{h})}\left(\frac{1}{h_j^2}\right)^{\alpha_{h}+1}\exp\left(-\frac{\beta_h}{h_j^2}\right),\qquad j=1,2,\dots,p, \\
\pi(b^2) &= \frac{(\beta_b)^{\alpha_b}}{\Gamma(\alpha_b)}\left(\frac{1}{b^2}\right)^{\alpha_b+1}\exp\left(-\frac{\beta_b}{b^2}\right),
\end{align*}
where $(\alpha_{\delta},\beta_{\delta})=(\alpha_h,\beta_h)=(\alpha_b,\beta_b)=(1.0, 0.05)$ are hyperparameters \citep[see for example,][]{Geweke10}. The selection of these hyperparameters is to assign more weights to a smaller value of squared bandwidth, in keeping consistent with the asymptotic results. The prior density with the chosen hyperparameter is shown in Figure~\ref{fig:prior}, along with two other possibilities considered in the sensitivity analysis given later in Table~\ref{tab:mcmctable}.

For discrete variables, denote $\pi(\lambda_s)$ as the prior of the discrete kernel bandwidth $\lambda_s$ for $s=1,2,\dots,q$. We assume the prior of $\lambda_s$ following a uniform distribution, given by
\begin{equation*}
\pi(\lambda_s)=\frac{1}{z_b-z_a},
\end{equation*}
where $\lambda_s\in [z_a,z_b]$. For binary variables, $z_a = 0$ and $z_b=0.5$; for ordered categorical variables, $z_a=0$ and $z_b=1.0$ \citep[see also][]{HR08}.

According to Bayes theorem, the posterior $\left(\delta^2, \bm{h}^2, \bm{\lambda}, b^2\right)$ can be expressed as (up to a normalising constant):
\begin{equation}
\pi\left(\delta^2,\bm{h}^2,\bm{\lambda},b^2|\bm{y}\right)\propto \widehat{L}\left(\bm{y}|\delta^2,\bm{h}^2,\bm{\lambda},b^2\right)\times \pi(\delta^2) \times \pi(\bm{h}^2)\times \pi(\bm{\lambda}) \times \pi(b^2),\label{eq:6}
\end{equation}
where $\widehat{L}(\bm{y}|\delta^2,\bm{h}^2,\bm{\lambda},b^2)$ is the same likelihood as in~\eqref{eq:5} but with squared bandwidths for those parameters that are estimated by the Gaussian kernel, and $\pi(\bm{h}^2) = \pi(h_1^2)\times \pi(h_2^2) \times \cdots \times \pi(h_p^2)$ and $\pi(\bm{\lambda})=\pi(\lambda_1)\times \pi(\lambda_2) \times \cdots \pi(\lambda_q)$ are independent priors. Since we assume that there is no correlation between the regression function and error term in~\eqref{eq:1}, the bandwidths of the regression function and error density are uncorrelated in~\eqref{eq:6}.

Based on~\eqref{eq:6}, we use the adaptive block random-walk Metropolis algorithm of \cite{GFS10} to sample $(\delta^2, \bm{h}^2, \bm{\lambda}, b^2)$. Throughout the paper, the burn-in period is the first 1,000 iterations, and the number of recorded iterations after the burn-in period is $10,000$ iterations. To measure the mixing performance of the sample paths, we consider the simulation inefficiency factor \citep[see also][]{KSC98,Shang13}. It can be interpreted as the sample mean from an sampler that draws iid observations from the posterior distribution.


In the kernel density estimation, it has been observed that the leave-one-out kernel estimator, such as~\eqref{eq:4}, can be heavily affected by the presence of extreme residuals \citep[see, for example,][]{Bowman84}. This may cause by the use of a global bandwidth. To rectify this problem, we consider the local bandwidth method studied by \cite{ZK11} and \cite{Shang13}. The idea of localised bandwidths is to assign small bandwidths to the observations in the high density region, while large bandwidths to the observations in the low density region. The localised kernel-form error density can be expressed by
\begin{equation*}
  \widehat{f}(\widehat{\varepsilon}_i;b,\tau_{\varepsilon}) = \frac{1}{n-1}\sum^n_{\substack{j=1\\j\neq i}}\frac{1}{b(1+\tau_{\varepsilon}|\widehat{\varepsilon}_j|)}\phi\left[\frac{\widehat{\varepsilon}_i-\widehat{\varepsilon}_j}{b(1+\tau_{\varepsilon}|\widehat{\varepsilon}_j|)}\right],
\end{equation*}
where $b(1+\tau_{\varepsilon}|\widehat{\varepsilon}_j|)$ is the bandwidth assigned to $\widehat{\varepsilon}_j$, for $j=1,2,\dots,n$, and the vector of parameters is now $\left(\delta,\bm{h},\bm{\lambda},b,\tau_{\varepsilon}\right)$. By setting the prior of $\tau_{\varepsilon}$ to be a uniform density on $[0,1]$, the adaptive random-walk Metropolis algorithm is used to sample these parameters.

\section{Simulation study} \label{sec:3}

The main goal of this section is to illustrate the proposed methodology through simulated data. One way to do that consists in comparing the true regression function with the estimated regression function, and comparing the true error density with the estimated error density. To measure the estimation accuracy between $m(\bm{z})$ and $\widehat{m}(\bm{z})$, we use the mean average squared error (MASE), given by
\begin{align}
\text{E}\int^b_a\left[m(\bm{z})-\widehat{m}_h(\bm{z})\right]^2d\bm{z}\approx \frac{1}{100}\sum_{\varsigma=1}^{100}\frac{b-a}{n}\sum_{i=1}^n \left[m^{\varsigma}(\bm{z_i})-\widehat{m}^{\varsigma}(\bm{z_i})\right]^2,\label{eq:31}
\end{align}
for $z \in [a,b]$, where $n$ represents the sample size, and $\varsigma$ represents one out of 100 replications considered.

To measure the discrepancy between $f(\varepsilon)$ and $\widehat{f}(\varepsilon)$, we calculate mean integrated squared error (MISE) defined as
\begin{align}
\text{E}\int^b_a\left[f(\varepsilon)-\widehat{f}(\varepsilon)\right]^2d\varepsilon &\approx \frac{1}{100}\sum^{100}_{\varsigma=1}\frac{b-a}{\kappa}\sum^{\kappa}_{i=1}\left[f^{\varsigma}(\varepsilon_i)-\widehat{f}^{\varsigma}(\varepsilon_i)\right]^2,\label{eq:32}
\end{align}
for $\varepsilon\in [a,b]$. For each replication, MISE can be approximated at $\kappa=1001$ grid points bounded between an interval, such as $[-5,5]$.

\begin{figure}[!ht]
  \centering
  \includegraphics[width=13.5cm]{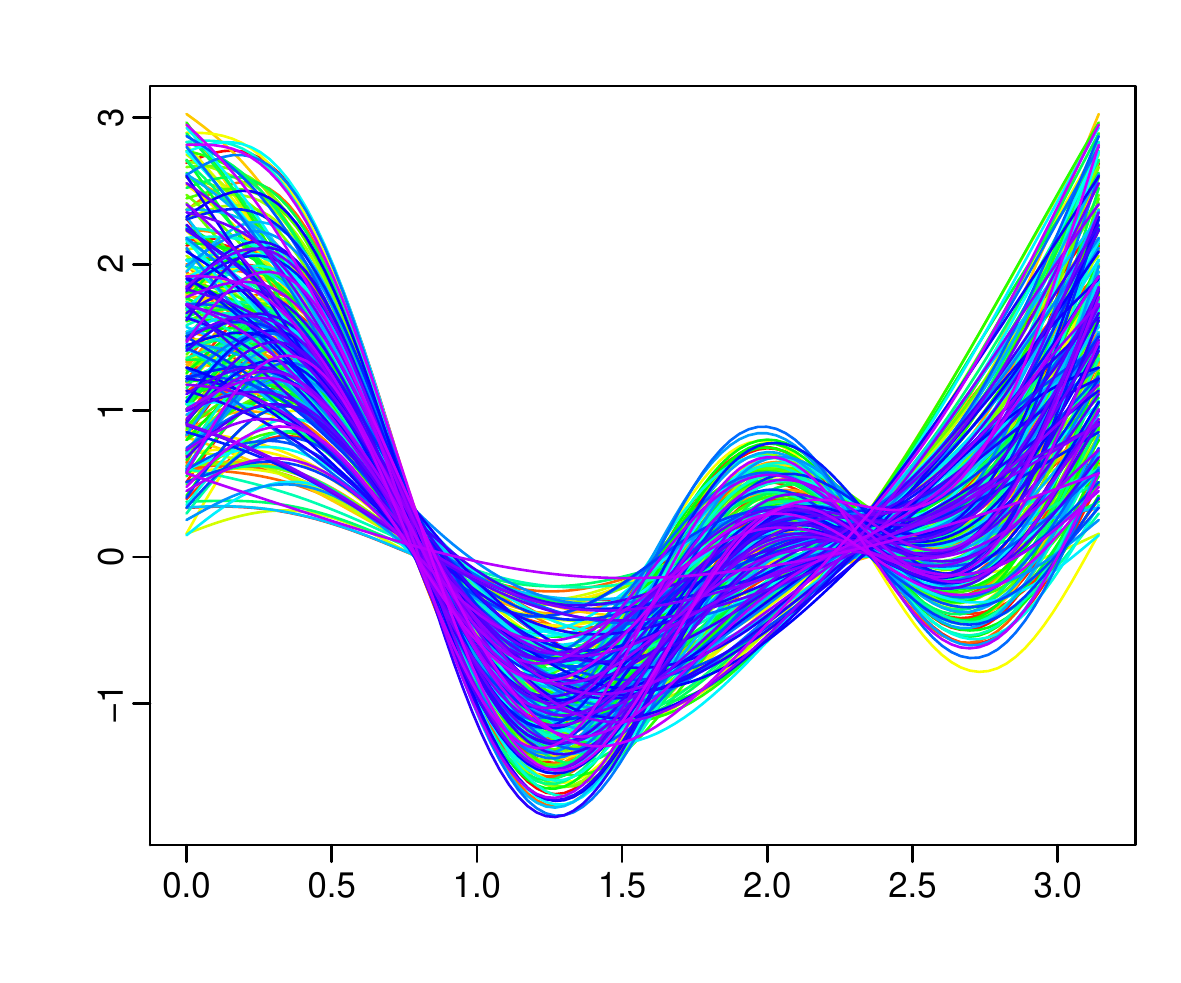}
  \caption{250 simulated curves.}\label{fig:1}
\end{figure}

\paragraph{Building the simulated samples.} We briefly describe the construction of the simulated data. First of all, we build simulated discretised curves
\begin{equation}
T_i(t_{v}) = a_i\cos(2t_v) + b_i\sin(4t_v) + c_i\left(t_v^2-\pi t_v+\frac{2}{9}\pi^2\right), \qquad i=1,2,\dots,n,\label{eq:11}
\end{equation}
where $0\leq t_1\leq t_2,\dots,\leq t_{100}\leq \pi$ are equispaced points, $a_i$, $b_i$, $c_i$ are independently drawn from a uniform distribution on $[0,1]$. The functional form of~\eqref{eq:11} is taken from \cite{FVV10}. Figure~\ref{fig:1} presents the simulated curves for one replication.

Once the curves are defined, one simulates a functional regression model to compute the responses:
\begin{itemize}
\item two regression functions are constructed as
\[ \left\{ \begin{array}{ll}
           \mbox{Model 1} & \mbox{$m\left(T_i,X_i^{\text{(c)}},X_i^{\text{(d)}}\right)=\int_0^{\pi} t cos(t) (T_i^{'}(t))^2 dt +\eta_i+\gamma_i,$}\\
            \mbox{Model 2} & \mbox{$m\left(T_i,\bm{X}_i^{\text{(c)}},\bm{X}_i^{\text{(d)}}\right)=\int_0^{\pi} t cos(t) (T_i^{'}(t))^2 dt +\eta_i+\omega_i+\gamma_i+\beta_i$},\end{array} \right. \]
where $T^{'}$ is the $1^{\text{st}}$ derivative of $T$, $\eta_i$ is a real-valued continuous variable that is simulated from a standard normal distribution, $\omega_i$ is a real-valued continuous variable that is simulated from an exponential distribution with rate parameter 1, $\gamma_i$ is a discrete-valued variable that is drawn from a Bernoulli distribution, and $\beta_i\in \{0,1,\dots,5\}$ with $P(\beta_i=l)=\frac{1}{6}$ for $l=0,1,\dots,5$.
\item generate $\varepsilon_1,\dots,\varepsilon_n$ from a trimodal distribution with functional form of $\frac{9}{20}N\left(-\frac{6}{5},(\frac{3}{5})^2\right) + \frac{9}{20}N\left(\frac{6}{5},(\frac{3}{5})^2\right) + \frac{1}{10}N\left(0,(\frac{1}{4})^2\right)$, for example. As shown in the supplement, we have also considered claw error density \citep[see also][Table 1]{MW92}.
\item compute the corresponding response $y_i = m\left(T_i,X_i^{\text{(c)}},X_i^{\text{(d)}}\right) + \varepsilon_i$ as in model 1.
\end{itemize}

\paragraph{Estimating the regression function.} We compute the discrepancy between $m(T_i, X_i^{\text{(c)}}, X_i^{\text{(d)}})$ and $\widehat{m}(T_i,X_i^{\text{(c)}},X_i^{\text{(d)}})$, for $i=1,2,\dots,n$. To do that, we use the following Monte-Carlo scheme:
\begin{enumerate}
\item build 100 replications $\left\{\left(y_i^{\varsigma},T_i^{\varsigma},X_i^{c,\varsigma},X_i^{d,\varsigma}\right)_{i=1,\dots,n}\right\}_{\varsigma=1,\dots,100}$.
\item compute the difference $\left\{\left[m\left(T_i,X_i^{\text{(c)}},X_i^{\text{(d)}}\right)-\widehat{m}^{\varsigma}\left(T_i,X_i^{\text{(c)}},X_i^{\text{(d)}}\right)\right]_{i=1,\dots,n}\right\}_{\varsigma=1,\dots,100}$, where $\widehat{m}^{\varsigma}\left(T_i,X_i^{\text{(c)}},X_i^{\text{(d)}}\right)$ represents the regression function estimated by the functional NW estimator with generalised product kernel for the $\varsigma^{\text{th}}$ replication. The Gaussian kernel function is used for function-valued and continuous real-valued predictors, while the  \citeauthor{AA76}'s \citeyearpar{AA76} kernel function is used for binary predictor and \citeauthor{LR07}'s \citeyearpar{LR07} kernel is used for categorical predictor.
\item For each replication, we calculate the average squared error of the regression function. Averaged over 100 replications, we use the MASE to access the estimation accuracy of the regression function.
\end{enumerate}

Table~\ref{tab:5} presents the MASE for the estimated regression function in the nonparametric functional regression model with mixed types of regressors. For both models, the functional cross validation method produces larger MASE than the Bayesian approach. Between a global bandwidth and localised bandwidths, we found that the former performs better in model 2 when there are more regressors; but the latter performs better in model 1 when there are less regressors. As $n$ increases, the estimation accuracy improves and the difference between the two Bayesian methods becomes smaller. For comparison, we also considered the same regression model without discrete-valued regressor. We found that the inclusion of discrete-valued regressor reduces MASE in both cases, more so for model 2 than model 1.

\begin{table}[!ht]
\begin{small}
\tabcolsep 0.04cm
  \begin{center}
    \begin{tabular}{lccccccccc@{}}\toprule
               &  \multicolumn{3}{c}{Functional cross validation} & \multicolumn{6}{c}{Bayesian method} \\
             & & & &   \multicolumn{3}{c}{Global bandwidth} & \multicolumn{3}{c}{Localised bandwidths} \\
        \backslashbox[2mm]{Type of regressor}{$n$} & 50 & 150 & 250 & 50       & 150      & 250      & 50       & 150      & 250      \\ \toprule
\multicolumn{10}{l}{function, continuous real-valued, discrete-valued}\\
model 1 & 2.196 & 1.512 & 1.307 & 2.115 & 1.201 & 1.038 & \textcolor{red}{2.084} & \textcolor{red}{1.194} & \textcolor{red}{1.031} \\
        & (0.723) & (0.246) & (0.167) & (0.742) & (0.213) & (0.148) & (\textcolor{blue}{0.710}) & (\textcolor{blue}{0.195}) & (\textcolor{blue}{0.143}) \\
model 2 & 2.201 & 1.614 & 1.341 & \textcolor{red}{1.812} & \textcolor{red}{1.372} & \textcolor{red}{1.213} & 3.035 & 2.374 & 2.211 \\
        & (2.165) & (0.271) & (0.211) & (\textcolor{blue}{0.568}) & (\textcolor{blue}{0.229}) & (\textcolor{blue}{0.145}) & (1.309) & (0.523) & (0.338) \\
\multicolumn{10}{l}{function, continuous real-valued}\\
model 1 & 2.304 & 1.570 & 1.402 & 2.440 & 1.335 & 1.139 & 2.417 & 1.335 & 1.140 \\
       & (0.716) & (0.257) & (0.198) & (0.863) & (0.246) & (0.159) & (0.873) & (0.228) & (0.155)  \\
model 2 & 3.201 & 3.078 & 2.806 & 3.936 & 2.895 & 2.758 & 3.988 & 3.257 & 3.110 \\
       & (1.240) & (0.705) & (0.539) & (1.629) & (0.703) & (0.572) & (1.657) & (0.689) & (0.513) \\\bottomrule
    \end{tabular}
    \caption{MASE comparison among the functional cross validation and two Bayesian bandwidth estimation methods with and without the inclusion of discrete-valued regressor for estimating the regression function. The red coloured text represents the minimal mean, while the blue coloured text represents the minimal sd. The number in parenthesis represents the sample standard deviation of the squared errors. The expression of MASE is shown in~\eqref{eq:31}.}\label{tab:5}
  \end{center}
  \end{small}
\end{table}

\paragraph{Estimating the error density} With a set of residuals, we apply a univariate kernel density estimator with a global bandwidth or local bandwidths. For $\varsigma=1,2,
\dots,100$, we compute the discrepancy in terms of integrated square error between $f^{\varsigma}(\varepsilon)$ and $\widehat{f}^{\varsigma}(\varepsilon)$, and obtain the overall discrepancy by averaging over 100 replications of discrepancy.

Table~\ref{tab:6} presents the MISE for the kernel-form error density with bandwidths estimated by the two Bayesian methods. There is an advantage in using the localised bandwidths over the global bandwidth, for all sample sizes considered.

\begin{table}[!ht]
\begin{small}
  \begin{center}
    \begin{tabular}{lccccccc@{}}\toprule
    						& \multicolumn{7}{c}{Bayesian method} \\
                                                & \multicolumn{3}{c}{Global bandwidth} &          & \multicolumn{3}{c}{Localised bandwidths} \\
        \backslashbox[2mm]{Type of regressor}{$n$} & 50       & 150      & 250      & & 50       & 150      & 250      \\ \toprule
\multicolumn{8}{l}{function, continuous real-valued, discrete-valued}\\
model 1 & 0.0896 & 0.0809 & 0.0599 & & \textcolor{red}{0.0233} & \textcolor{red}{0.0088} & \textcolor{red}{0.0057} \\
        & (0.0513) & (0.0482) & (0.0231) & & (\textcolor{blue}{0.0199}) & (\textcolor{blue}{0.0099}) & (\textcolor{blue}{0.0043}) \\
model 2 & 0.2839 & 0.1503 & 0.1217 & & \textcolor{red}{0.0287} & \textcolor{red}{0.0204} & \textcolor{red}{0.0184} \\
        & (0.5101) & (0.1139) & (0.0585) & & (\textcolor{blue}{0.0210}) & (0.0148) & (0.0089) \\
\multicolumn{8}{l}{function, continuous real-valued}\\
model 1 & 0.0522 & 0.0432 & 0.0330 & & 0.0545 & 0.0494 & 0.0414 \\
        & (0.0458) & (0.0279) & (0.0129) & & (0.0467) & (0.0248) & (0.0146) \\
model 2 & 0.0672 & 0.1286 & 0.1266 & & 0.0619 & 0.0442 & 0.0424 \\
        & (0.0636) & (0.0253) & (0.0256) & & (0.0694) & (\textcolor{blue}{0.0085}) & (\textcolor{blue}{0.0051}) \\\bottomrule
    \end{tabular}
  \caption{MISE comparison between the two Bayesian bandwidth estimation methods with and without the inclusion of discrete-valued regressor for estimating the error density. The number in parenthesis represents the sample standard deviation of the squared errors. The expression of MISE is shown in~\eqref{eq:32}.}\label{tab:6}
  \end{center}
\end{small}
\end{table}

\begin{figure}[!htbp]
  \centering
  {\includegraphics[width=6.9cm]{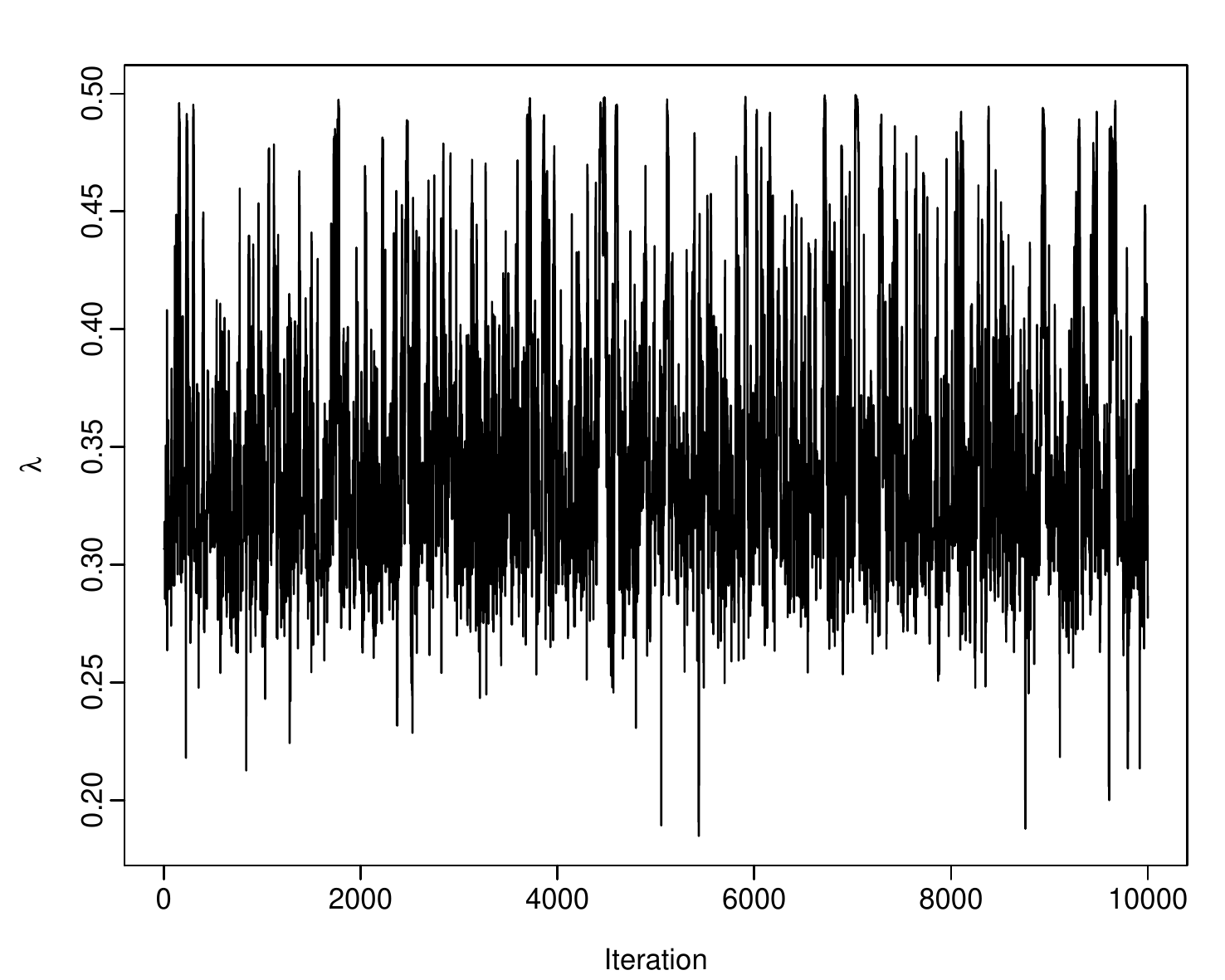}}\quad
  {\includegraphics[width=6.9cm]{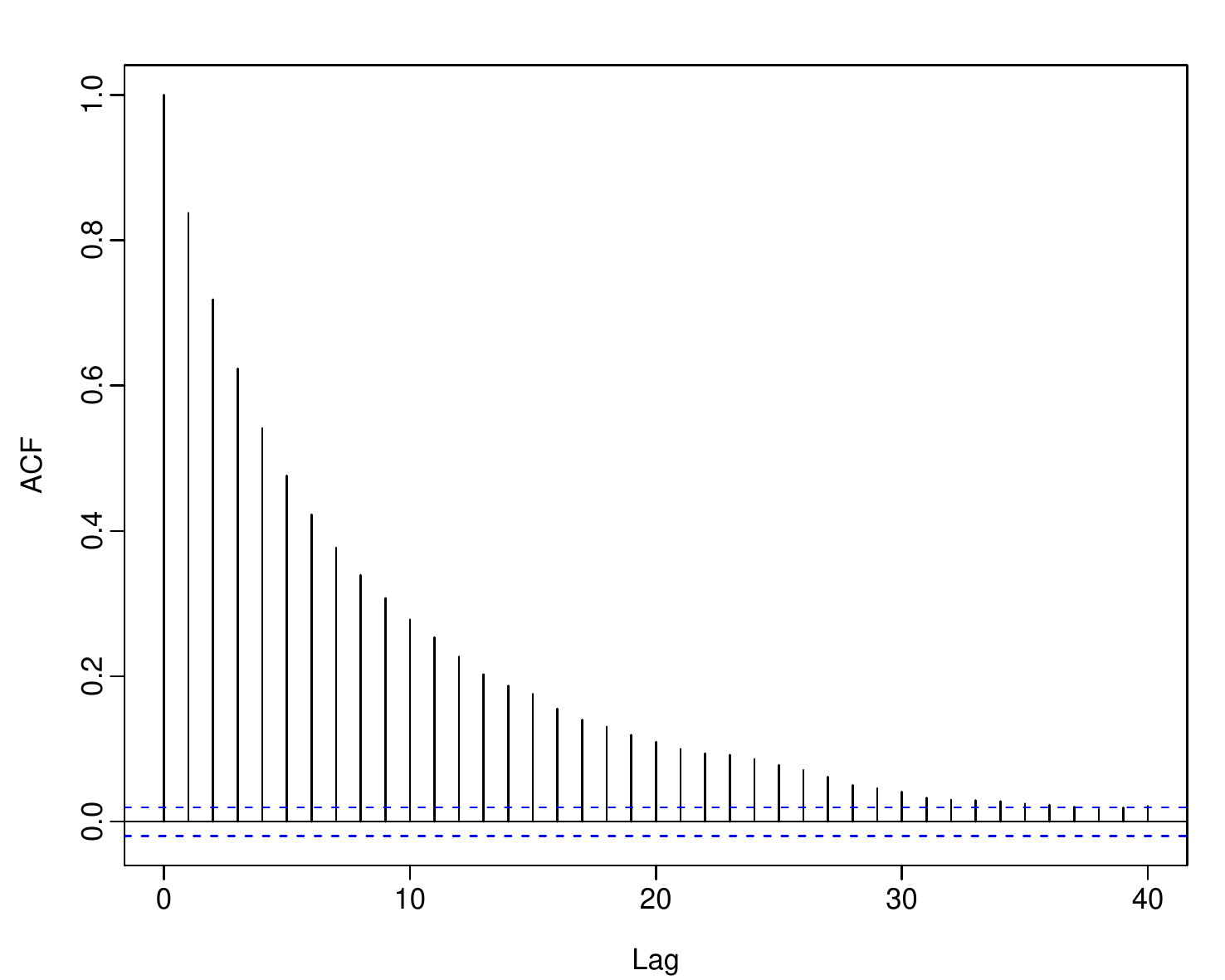}}\\
  {\includegraphics[width=6.9cm]{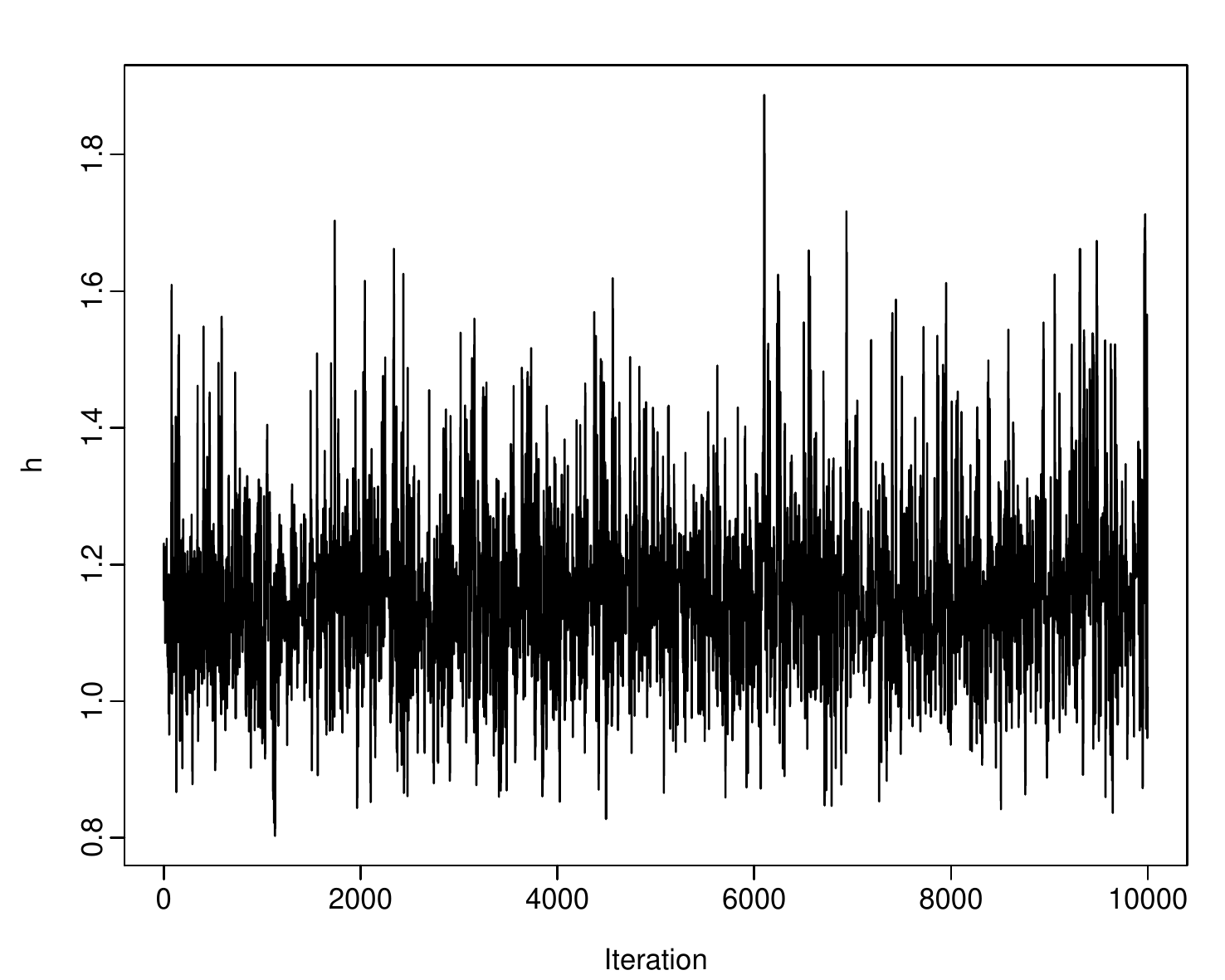}} \quad
  {\includegraphics[width=6.9cm]{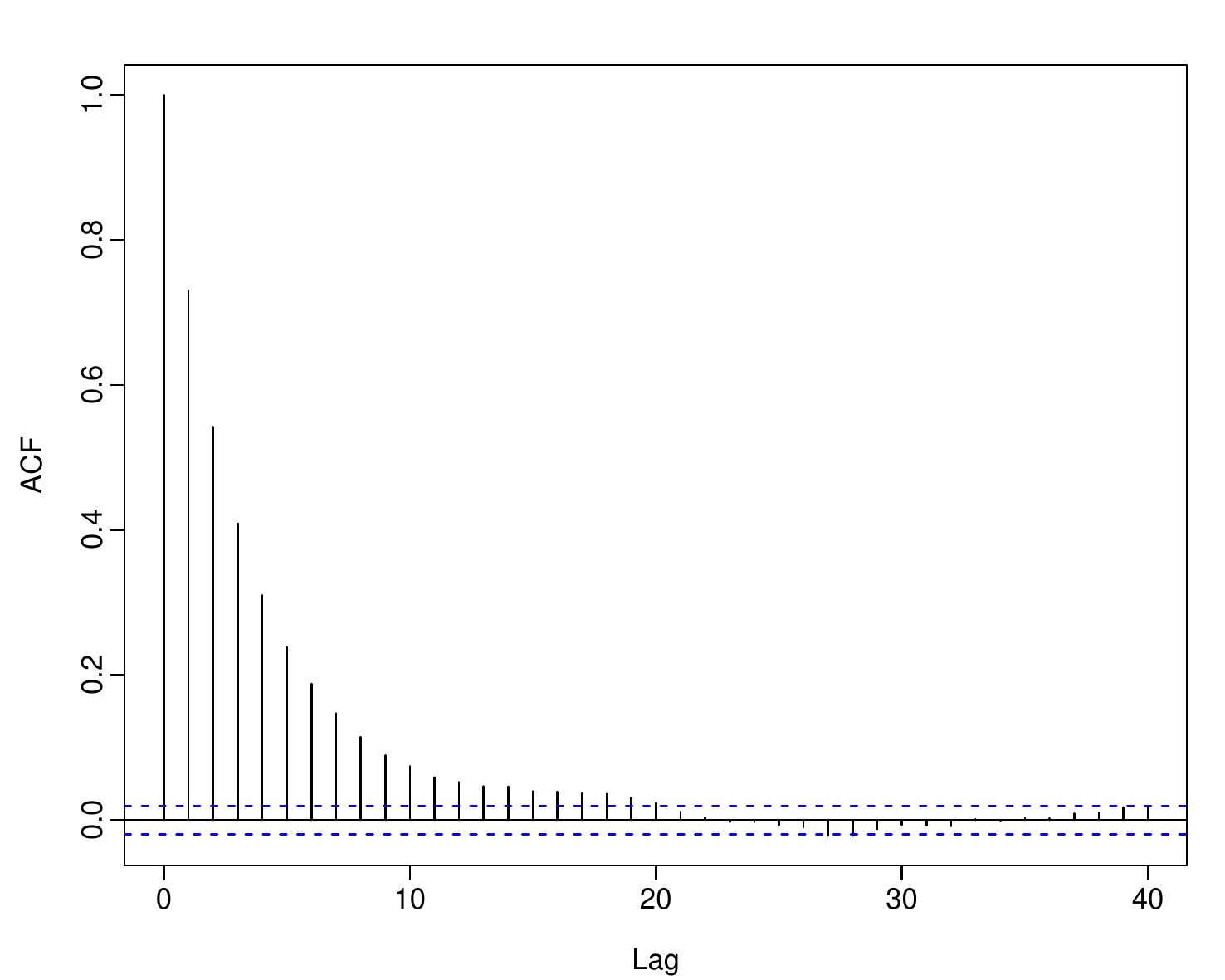}}\\
  {\includegraphics[width=6.9cm]{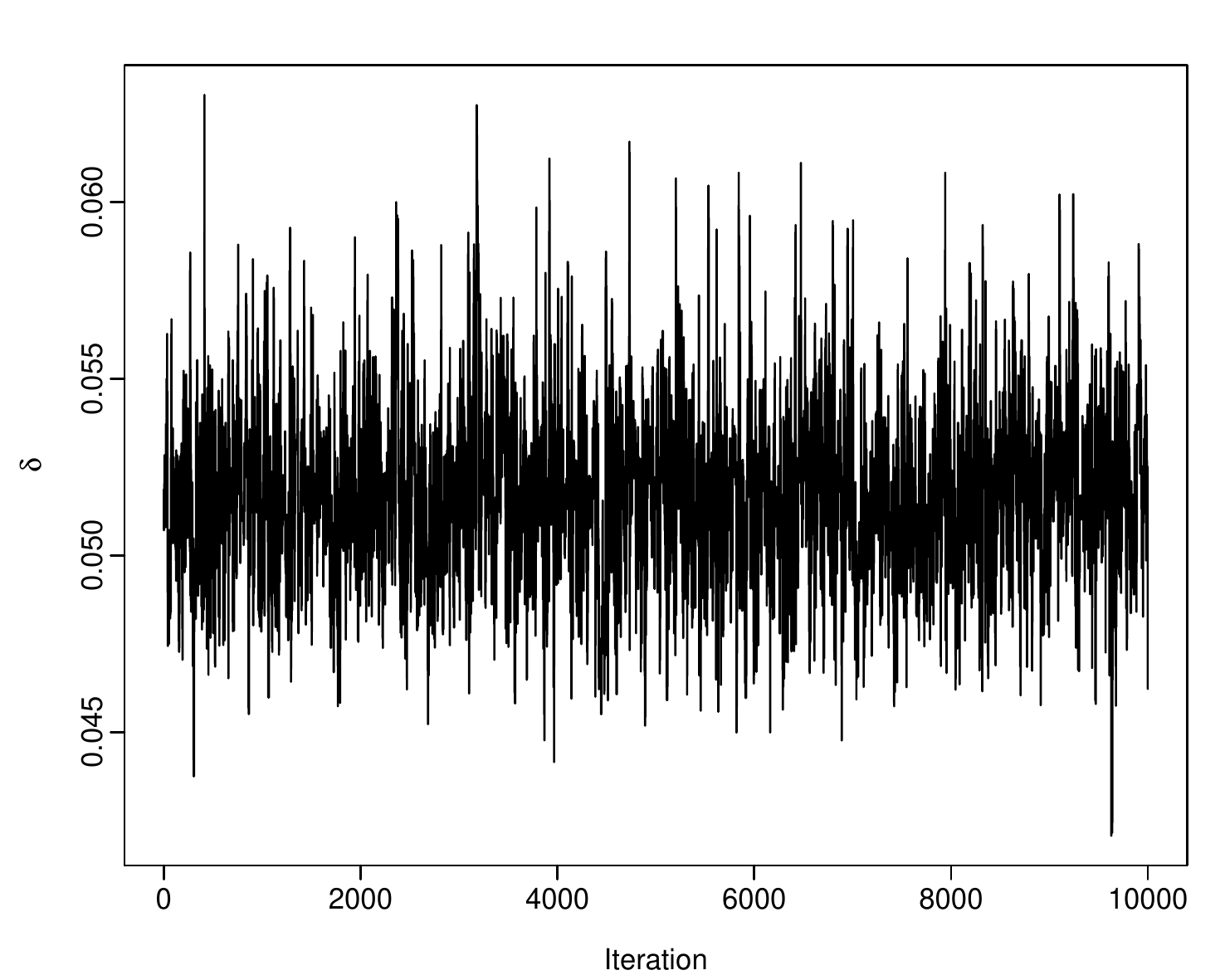}} \quad
  {\includegraphics[width=6.9cm]{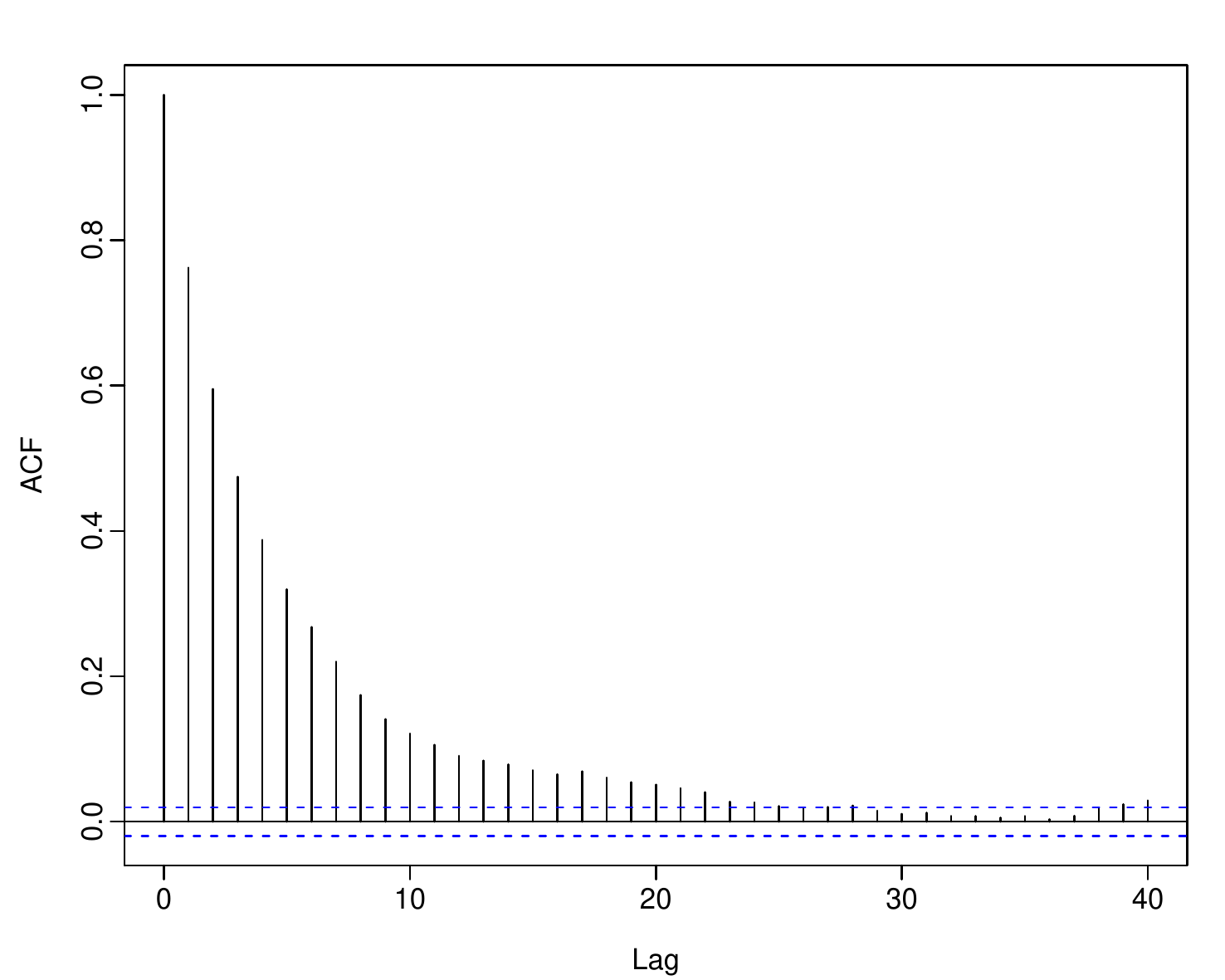}}\\
  {\includegraphics[width=6.9cm]{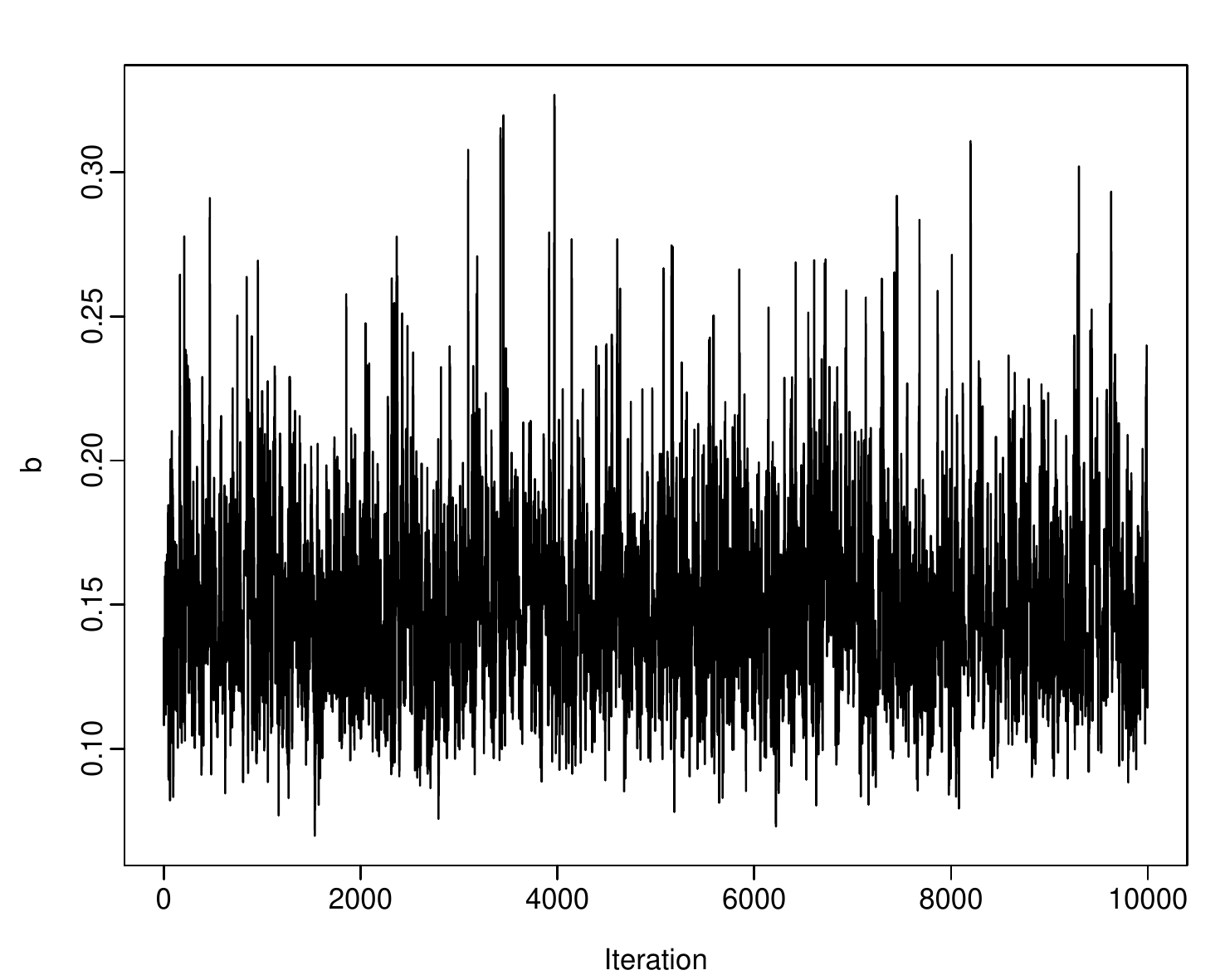}} \quad
  {\includegraphics[width=6.9cm]{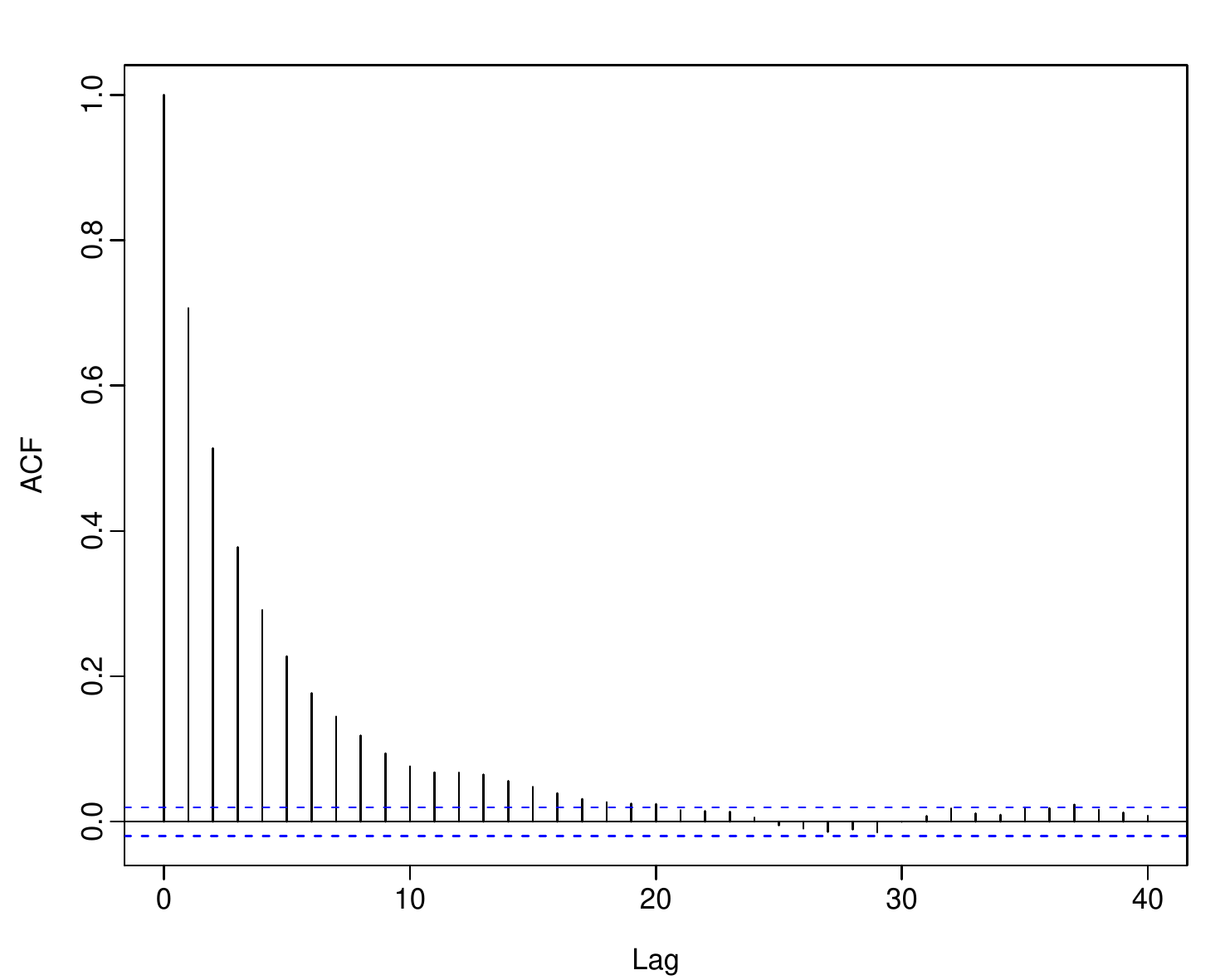}}
  \caption{On the left panel, trace plots of the bandwidths for the discrete regressor ($\lambda$), continuous regressor ($h$), functional regressor ($\delta$), and residual ($b$). On the right panel, ACF plots of the bandwidths.}\label{fig:mcmcplot}
\end{figure}

\paragraph{Diagnostic check of Markov chains} As a demonstration with one replication, we plot the un-thinned sample paths of all the parameters in model 1 on the left panel of Figure~\ref{fig:mcmcplot}, and the ACFs of these sample paths on the right panel of Figure~\ref{fig:mcmcplot}. These plots show that the sample paths are reasonably well mixed. Table~\ref{tab:mcmctable} summarises the ergodic averages, 95\% Bayesian credible intervals (CIs), total SE, batch-mean SE, and SIF values.

\begin{table}[!ht]
\tabcolsep=0.48cm
  \begin{center}
    \begin{tabular}{@{}lccccr@{}}
    \multicolumn{3}{l}{\hspace{-.24in} Prior density: IG($\alpha=1, \beta=0.05$)} & & & \\\cline{1-3}
    Parameter & Mean & Bayesian CIs & SE & Batch-mean SE & SIF \\\toprule
    $\lambda$                                                     & 0.3716   & (0.1935, 0.4937)  & 0.0840 & 0.2630       & 9.81   \\
    $h$	                                                          & 0.8119   & (0.6511, 1.0322)  & 0.0985 & 0.2263       & 5.28	  \\
    $\delta$	                                                  & 0.7249   & (0.6509, 0.8131)  & 0.0422 & 0.0951       & 5.07	  \\
    $b$                                                           & 0.2001   & (0.0988, 0.3283)  & 0.0638 & 0.1956       & 9.40	  \\\toprule
    \multicolumn{3}{l}{\hspace{-.24in} Prior density: IG($\alpha=5, \beta=0.25$)} &          &                   &        \\ \cline{1-3}
    $\lambda$                                                     & 0.3493   & (0.1823, 0.4806)  & 0.0881 & 0.2173       & 6.08   \\
    $h$	                                                          & 0.7540   & (0.5995, 0.9285)  & 0.0794 & 0.1747       & 	4.84  \\
    $\delta$                                                      & 0.7041   & 	(0.6090, 0.7840) & 0.0432 & 0.1093       & 6.39   \\
    $b$	                                                          & 0.2200   & (0.1516, 0.3011)  & 0.0392 & 0.1067       & 	7.41  \\\toprule
    \multicolumn{3}{l}{\hspace{-.24in} Prior density: Cauchy}                     &          &                   &        \\ \cline{1-3}
    $\lambda$                                                     & 0.3632   & (0.2118, 0.4795)  & 0.0742 & 0.2305       & 9.65   \\
    $h$	                                                          & 0.8393   & (0.6650, 1.0830)  & 0.1059 & 0.2765       & 6.82   \\
    $\delta$	                                                  & 0.7316   & (0.6564, 0.8296)  & 0.0417 & 0.1003       & 5.79	  \\
    $b$                                                           & 0.1925   & (0.0542, 0.3590)  & 0.0875 & 0.2987       & 11.66	 \\\bottomrule
    \end{tabular}
    \caption{MCMC results of the bandwidth estimation under different prior densities with trimodal error density and $n=250$.}\label{tab:mcmctable}
  \end{center}
\end{table}

By using the coda package \citep{PBC+06} in {\Rx{}} language \citep{Team12}, we further checked the convergence of Markov chain with \possessivecite{Geweke92} convergence diagnostic test and \possessivecite{HW83} convergence diagnostic test. Our Markov chains pass both tests for the 100 replications.

\subsection{Regression models having irrelevant regressors}

Sometimes, not all the regressors in $\bm{z}_i=\left(T_i, \bm{X}_i^{\text{(c)}}, \bm{X}_i^{\text{(d)}}\right)$ are relevant \citep[see][for the case of nonparametric multivariate regression model]{HLR07, MR12}. Without loss of generality, we assume that some of the continuous and discrete regressors are irrelevant and set up our simulation study using the previous two models. Suppose the true model contains only one function-valued, one real-valued continuous and one discrete-valued regressors. However, we observe one function-valued, two real-valued continuous and two discrete-valued regressors. As pointed out in \cite{HLR07} and also shown in Table~\ref{tab:irrelevant}, the irrelevant discrete regressor is smoothed out when its bandwidth $\lambda$ takes on its upper bound value, while the irrelevant continuous regressor is effectively smoothed out when its bandwidth exceeds just a few standard deviation of the data.

\begin{table}[!htb]
\centering
\begin{tabular}{@{}lcccc@{}}\toprule
$n$ & 50 & 150 & 250 & 1000 \\\toprule
$\lambda_1$ & 0.128 [0.059, 0.175] & 0.329 [0.280, 0.386] & 0.330 [0.231, 0.410] & 0.275 [0.244, 0.305]  \\
$\lambda_2$ & 0.895 [0.823, 0.990] & 0.952 [0.860, 0.994] & 0.987 [0.955, 0.998] & 0.969 [0.916, 0.995] \\
$h_1$	      & 0.881 [0.608, 2.529] & 0.742 [0.669, 0.830] & 0.894 [0.766, 1.058] & 0.768 [0.716, 0.822] \\
$h_2$	      & 1.965 [0.997, 2.625] & 3.752 [1.977, 10.866] & 2.334 [1.478, 5.852] & 5.394 [3.145, 14.808] \\
$\delta$	      & 0.072 [0.068, 0.087] & 0.067 [0.064, 0.074] & 0.050 [0.047, 0.053] & 0.044 [0.042, 0.045]\\\bottomrule
\end{tabular}
\caption{Summary of Bayesian bandwidths (median, 10th percentile, 90th percentile of the bandwidths) for the functional NW estimator with mixed types of regressors. The standard deviation of the first continuous regressor is 1.0327, and the standard deviation of the second continuous regressor is 0.9758.}\label{tab:irrelevant}
\end{table}

The asymptotic results of irrelevant regressors in the nonparametric multivariate regression have been proven in \cite{HLR07}. We verified this phenomenon in the nonparametric functional regression with bandwidths selected by the Bayesian approach.

\section{Application to food quality control}\label{sec:4}

\begin{figure}[!ht]
  \centering
  \subfloat[Spectrometric curves: curves with fat content $<20\%$ (red lines) and curves with fat content $>20\%$ (blue lines).]
  {\includegraphics[width=7.3cm]{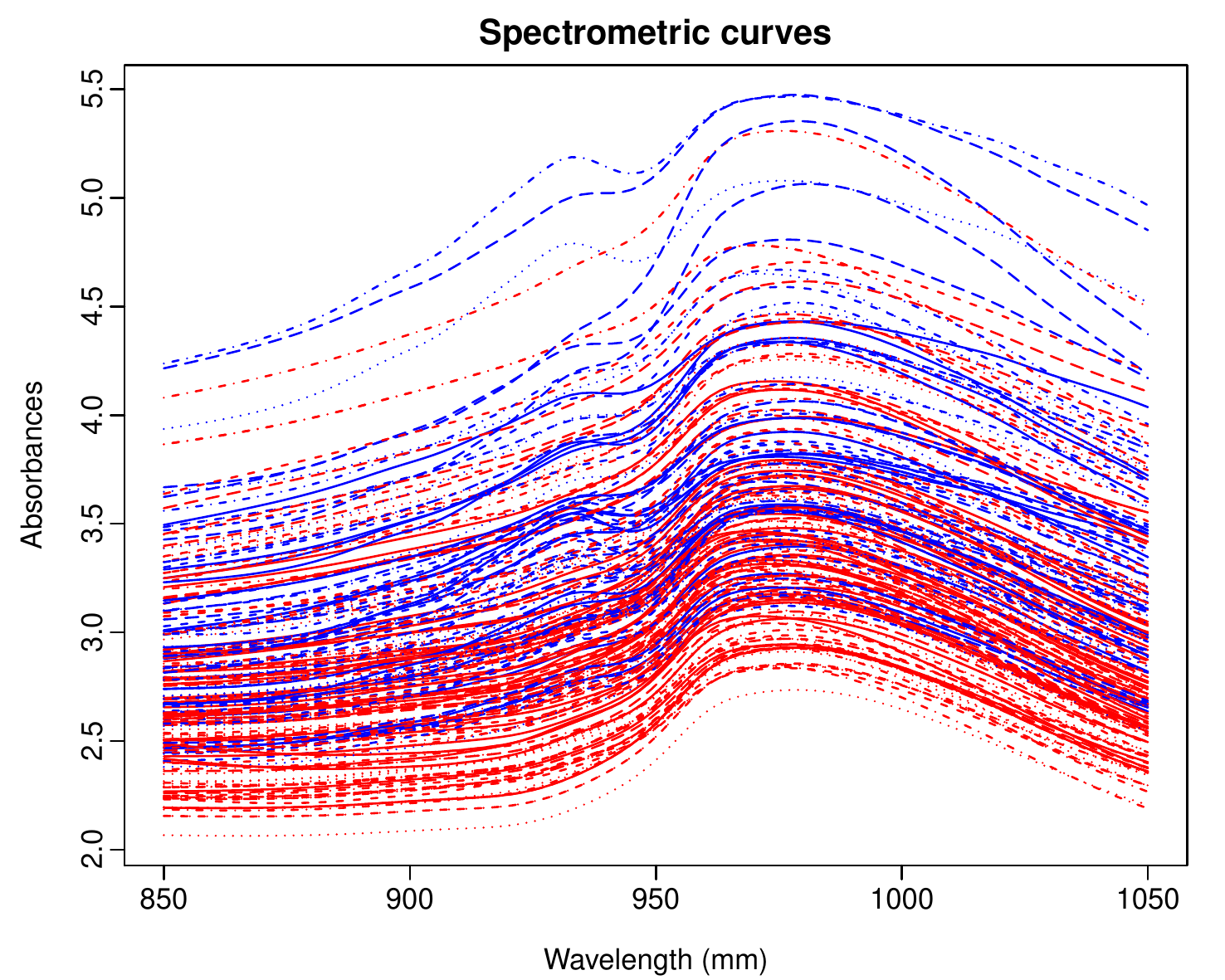}}
  \quad
  \subfloat[Corresponding spectrometric curves after first-order differencing.]
  {\includegraphics[width=7.3cm]{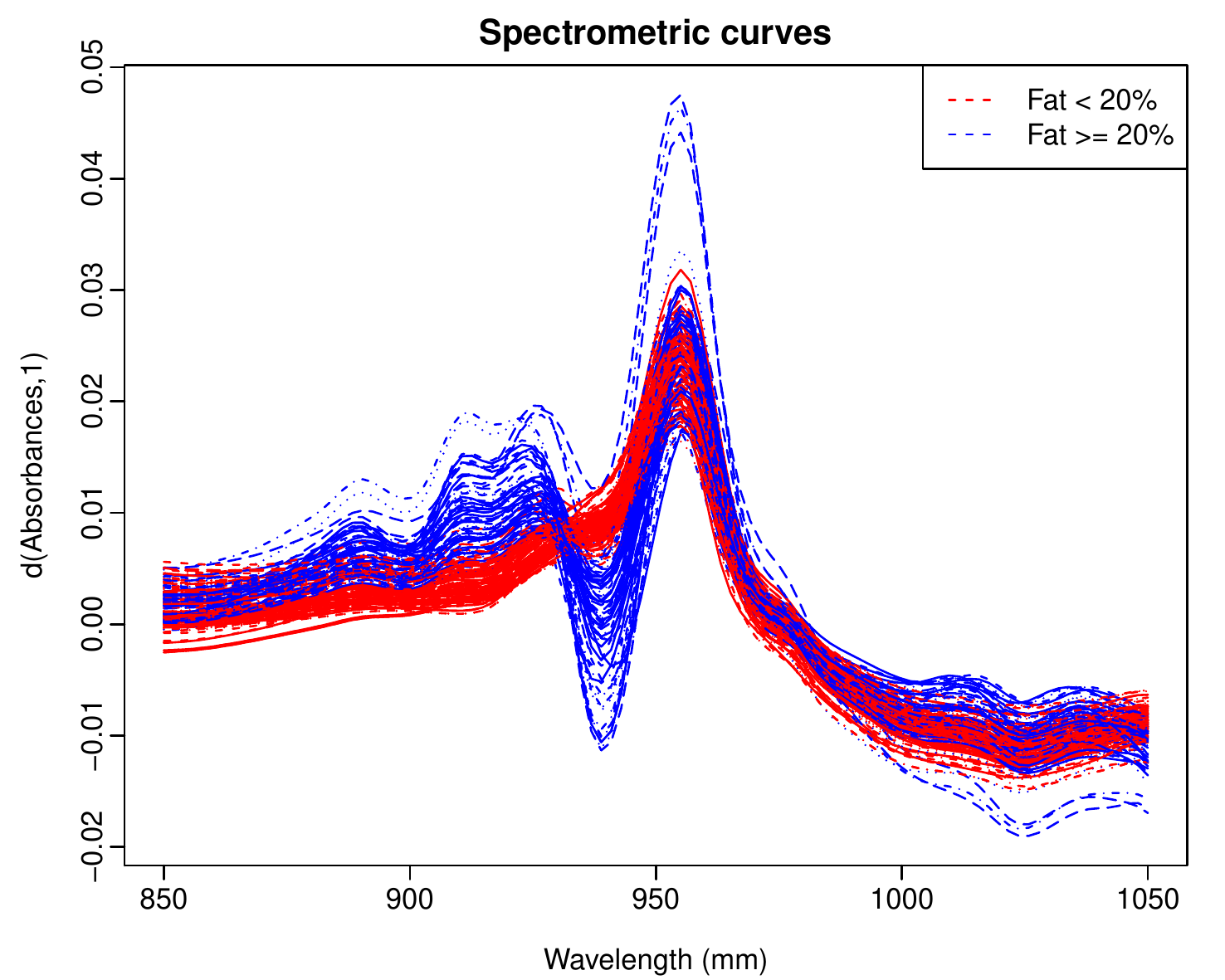}}
  \caption{Graphical displays of spectrometric curves.}\label{fig:2}
\end{figure}

Our second dataset focuses on the prediction of the fat content of meat samples based on near-infrared (NIR) absorbance spectra. These data were obtained from \url{http://lib.stat.cmu.edu/datasets/tecator}, and have been studied by \cite{FV06}, \cite{AV06}, among many others. Each food sample contains finely chopped pure meat with different percentages of the fat, protein and moisture contents. For each unit $i$ (among 215 pieces of finely chopped meat), we observe one spectrometric curve, denoted by $T_i$, which corresponds to the absorbance measured at a grid of 100 wavelengths (i.e., $T_i = (T_i(t_1),\dots,T_i(t_{100}))$. For each unit $i$, we also observe its fat, protein and moisture contents $\bm{X}\in R^3$, obtained by analytical chemical processing. As noted by \cite{FV03} and \cite{FF12}, the spectrometric curves can be split into two groups, based on if the fat content is below 20\%. Graphical displays of original spectrometric curves and their first derivative are shown in Figure~\ref{fig:2}.

Given a set of spectrometric curves, a classification algorithm can be used to find its group member. Given a new spectrometric curve $T_{\text{new}}$, we can allocate this new curve to its corresponding group member, denoted by $X_{\text{new}}^{d}$. Also, we may observe new measurements of protein and moisture contents, denoted by $X^{\text{(c)}}_{\text{new},1}$ and $X^{\text{(c)}}_{\text{new},2}$, which may help in the prediction of the fat content, denoted by $y_{\text{new}}$.

In order to assess the out-of-sample point forecast accuracy of the nonparametric functional estimator, we split the original sample into two sub-samples \citep[see also][p.105]{FV06}. The first one is called learning sample, which contains the first 160 units $\left\{y_i, T_i, X_i^{\text{(d)}}, X_{i1}^{\text{(c)}}, X_{i2}^{\text{(c)}}\right\}_{i=1,\dots,160}$. The second one is called testing sample, which contains the last 55 units $\left\{y_j, T_j, X_j^{\text{(d)}}, X_{j1}^{\text{(c)}}, X_{j2}^{\text{(c)}}\right\}_{j=161,\dots,215}$. The learning sample allows us to build the regression model with the optimal bandwidths, where the learning sample is used. The testing sample allows us to evaluate the prediction accuracy, where the testing sample is used.

To measure the prediction accuracy, we consider the mean square forecast error (MSFE) and mean absolute forecast error (MAFE). These are expressed as: MSFE = $\frac{1}{55}\sum^{215}_{\varpi=161}(y_{\varpi}-\widehat{y}_{\varpi})^2$ and MAFE = $\frac{1}{55}\sum^{215}_{\varpi=161}|y_{\varpi}-\widehat{y}_{\varpi}|$. The corresponding values of MSFE and MAFE for the functional cross validation and the two Bayesian bandwidth estimation methods are shown in Table~\ref{tab:1}. As a result, there is an improvement in prediction accuracy for the functional estimator with localised bandwidths than a global bandwidth.
\begin{table}[!ht]
\tabcolsep .13in
  \centering
  \begin{tabular}{@{}lcccc@{}}\toprule
   Method  &  MSFE & MAFE & Log marginal likelihood & Coverage probability \\\toprule
   \multicolumn{4}{l}{\hspace{-.18in} Functional cross validation}  & \\
   				& 1.5107 & 0.8738 \\
   			    & (2.7580) & (0.8723) \\
			    \\
   \multicolumn{4}{l}{\hspace{-.18in} Bayesian method with global bandwidth} \\
IG(1, 0.05)  &   1.5803 & 0.8718  & -227.04  & 0.78 \\
	      &   (3.0387)	& (1.2638)  \\
IG(5, 0.25)  &   1.6278  & 0.8862 & -233.57  & 0.76 \\
	       &  (3.1131)	& (1.2823)  \\
Cauchy  &   1.5853 &   0.8700  & -225.91  &  0.78 \\
	       &   (3.0758) & (1.2660) \\ 	\\
\multicolumn{4}{l}{\hspace{-.18in} Bayesian method with local bandwidth} \\
IG(1, 0.05) & 1.5015  & 0.8425 & -228.88 & 0.91 \\
	      & (2.9014) & (0.8980) \\
IG(5, 0.25) & 1.6979  & 0.8880 & -234.17 & 0.84 \\
	      & ( 3.4346) & (0.9624) \\ 	
Cauchy & 1.5039    & 0.8475    & -225.39 &  0.93 \\
	      & (2.9702) & (0.8945) \\\bottomrule
    \end{tabular}
  \caption{Out-of-sample MSFE and MAFE, log marginal likelihood and empirical coverage probability for the functional cross validation and Bayesian bandwidth estimation methods for forecasting fat content. The number in parenthesis represents the sample standard deviation of the squared or absolute errors.}\label{tab:1}
\end{table}

The model evidence, as measured by the log marginal likelihood, is an important quantity in the comparison of statistical models under the Bayesian paradigm. It has received numerous attention in the Bayesian statistics literature, such as the Laplace's method \citep{TK86}, harmonic mean estimator \citep{NR94}, generic method based on MCMC output \citep{Chib95, CJ01}, nested sampling \citep{Skilling06}, power posterior \citep{FP08}, to name only a few. \cite{FW12} provided a comparison of these methods, and found that the generic method based on MCMC output has good estimation accuracy, as well as fast computational speed. It is this method we consider here to compare the evidence from different prior distributions. From Table~\ref{tab:1}, we find that the three prior densities have similar marginal likelihoods, with the Cauchy prior density as the favour one (largest log marginal likelihood and best empirical coverage probability).

Because the original functional data are iid, we sample with replacement to obtain 100 replications of bootstrapped samples. While 160 pairs of observations within each replication are used for estimation, the remaining ones are used for forecast evaluation. Furthermore, we consider a large set of functional regression models collected in \cite{GS14} and \cite{FV11}. These models include: (1) REML-based functional linear model with a locally adaptive penalty \citep{CFS03}; (2) penalised functional regression \citep{GBC+11}; (3) functional principal component regression on first few functional principal components \citep{RO07}; (4) linear model on first $K$ functional principal components, where optimal $K$ is estimated by 20-fold bootstrap \citep{RS05}; (5) REML-based single-index signal regression with locally adaptive penalty \citep{Wood11}; (6) cross validation based single-index signal regression \citep{ELM09}; (7) penalised partial least squares \citep{KBT08}; (8) LASSO penalised linear model on first few functional principal components \citep{FHT10}; (9) nonparametric functional regression  with Nadaraya-Watson estimator for estimating conditional mean \citep{FV06}; (10) nonparametric functional regression with Nadaraya-Watson estimator for estimating conditional median \citep{LLS11}; (11) nonparametric functional regression with Nadaraya-Watson estimator for estimating conditional mode \citep{FLV05}; (12) nonparametric functional regression with $k$ nearest neighbour estimator \citep{BFV09}; (13) nonparametric functional regression with most-predictive design points \citep{FHV10}; (14) nonparametric functional regression with mixed types of regressors. 

\begin{figure}[!ht]
\centering
\subfloat[Input variable is a set of original functional curves.]
{\includegraphics[width=13.4cm]{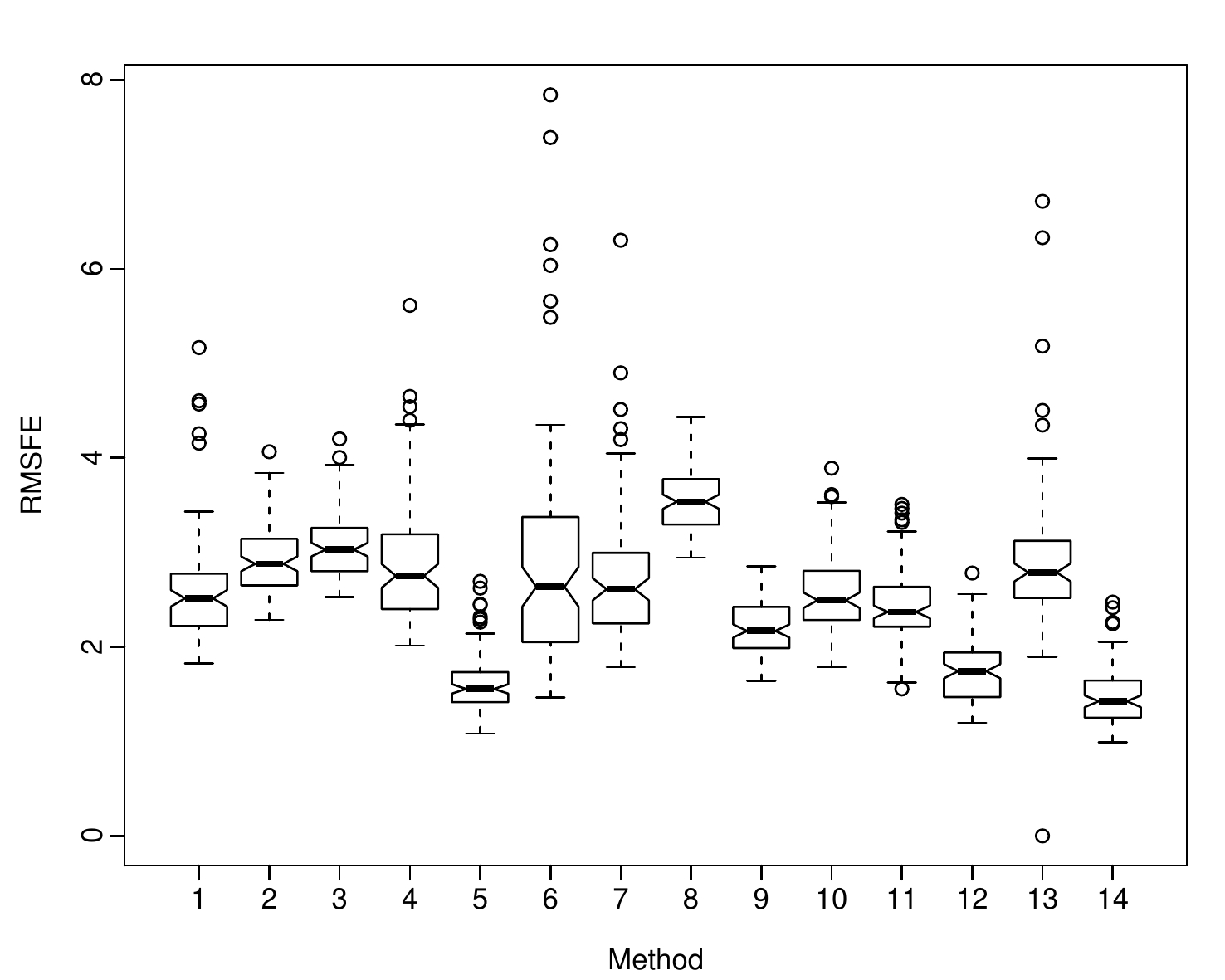}\label{fig:spec_1}}
\\
\subfloat[Input variable is the second derivative of original functional curves.]
{\includegraphics[width=13.4cm]{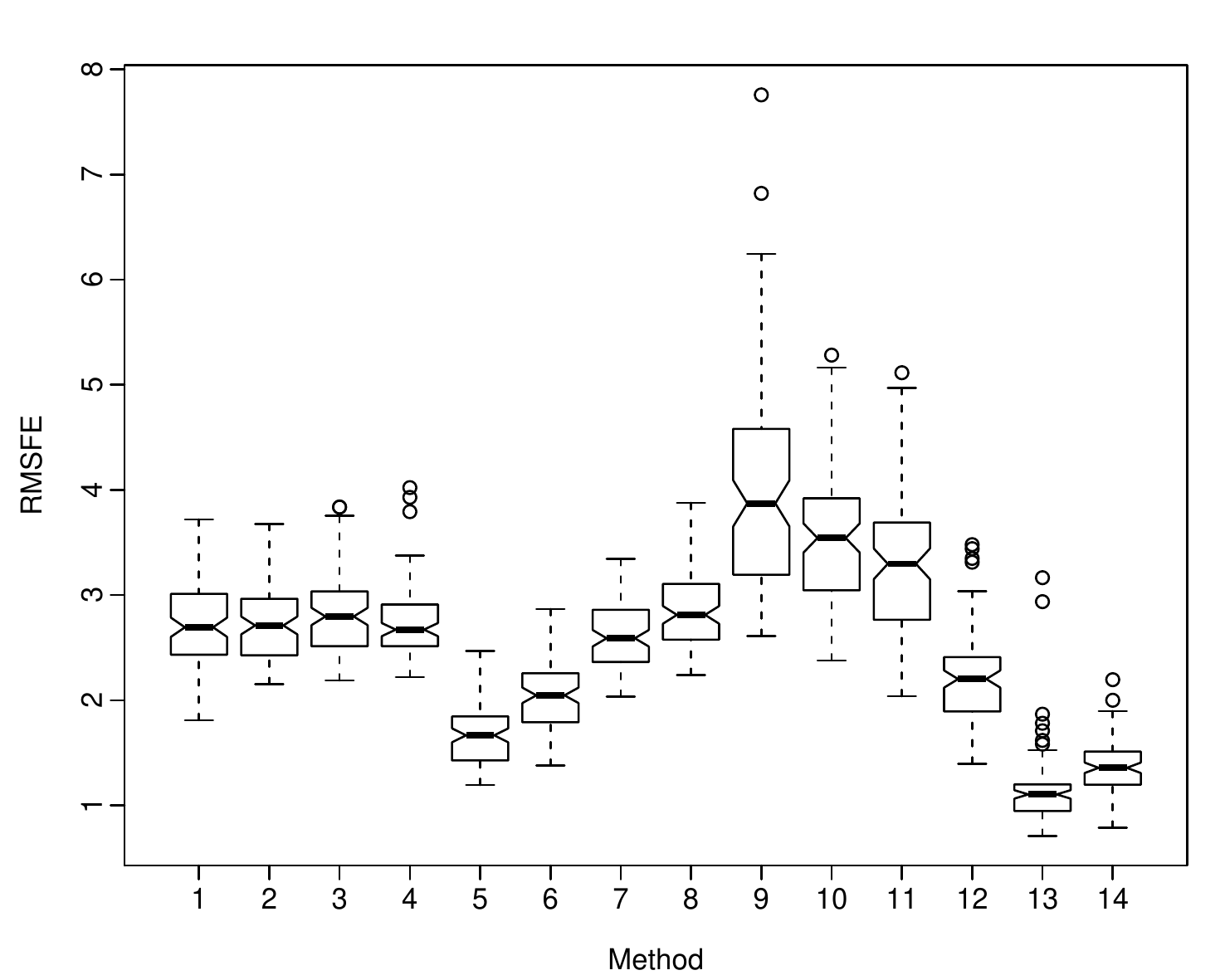}\label{fig:spec_2}}
\caption{Boxplots of forecast accuracy among several functional regression models. Refer to Section~\ref{sec:4} for details of the 14 methods.}\label{fig:spec_3}
\end{figure}

In Figure~\ref{fig:spec_1}, we investigated the forecast accuracy when the predictor is a set of original functional curves. The proposed nonparametric functional regression with mixed types of regressors gives the smallest root mean square forecast error (RMSFE) among all methods considered. In Figure~\ref{fig:spec_2}, we investigated the forecast accuracy when the predictor is second-derivative of original functional curves. The nonparametric functional regression with most-predictive design points performs the best in all, followed by the nonparametric functional regression with mixed types of regressors. Since the input variable affects the forecast accuracy of a method, it remains as a future research topic to investigate optimal transformation of the input variable in order to achieve maximum forecast accuracy.

\begin{figure}[!ht]
\centering
\includegraphics[width=\textwidth]{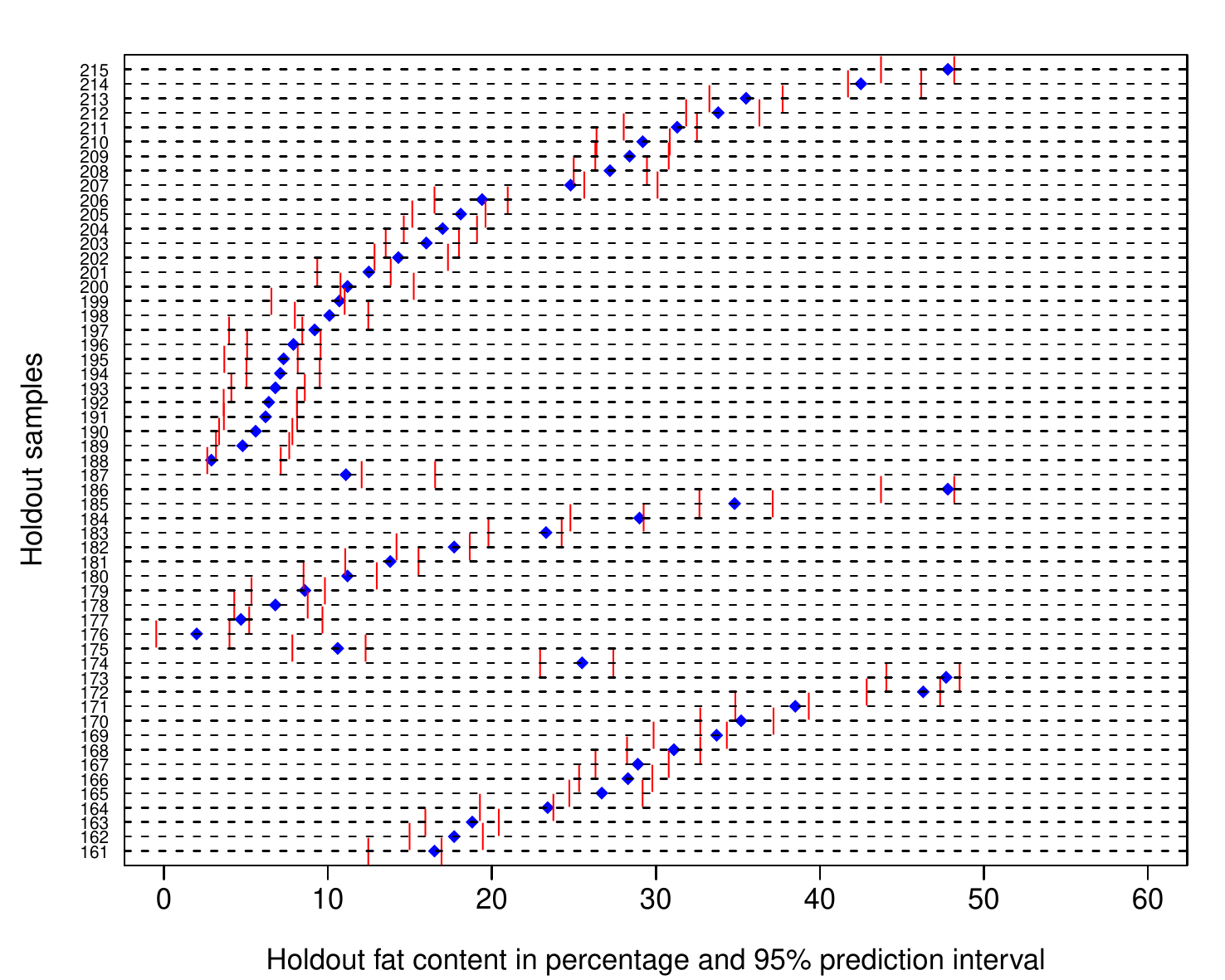}
\caption{Plot of predicted fat contents in percentage and the 95\% pointwise prediction intervals. The point forecasts of the fat content are shown as solid blue dots, while the 95\% prediction intervals are shown as vertical red bars. The localised bandwidths of error density are estimated by the Bayesian method with the Cauchy prior distribution. The empirical coverage probability is 93\%.}\label{fig:3}
\end{figure}

With the Bayesian approach, we can also compute the prediction interval nonparametrically. To this end, we first compute the cumulative density function (cdf) of the error distribution, over a set of grid points within a range, say -10 and 10. Then, we take the inverse of the cdf and find two grid points that are closest to the 2.5\% and 97.5\% quantiles. The 95\% prediction interval of the holdout samples is obtained by adding the two grid points to the point forecasts. For instance, the point forecasts of the fat content are shown in solid blue dots, while the 95\% pointwise prediction intervals are shown as vertical red bars in Figure~\ref{fig:3}. 

\section{Conclusion and future research}

We propose a Bayesian approach to estimate optimal bandwidths in a nonparametric functional regression model that admits mixed types of regressors with homoscedastic errors and unknown error density. Since a closed form expression for our bandwidth estimator does not exist, so establishing the mathematical properties of the bandwidth estimator has been difficult. This is true even in terms of the standard class of asymptotic statistics. However, we have developed an approximate solution to the bandwidth estimator through Markov chain Monte Carlo. As a byproduct, marginal likelihood can be used to determine the optimal choice of prior density for the bandwidths. 

Through a simulation study, the Bayesian approach provides a way to simultaneously estimate the bandwidths in the functional NW estimator and kernel-form error density. Illustrated by a spectroscopy data set, the Bayesian bandwidth estimation approach allows the nonparametric construction of prediction interval for measuring the prediction uncertainty. To the best of our knowledge, the proposed method makes the first attempt in modeling and forecasting the scalar real-valued response variable based on mixed types of regressors.

There are many ways in which the present paper can be extended, and we briefly mention five at this point:
\begin{compactenum}
  \item  Consider other functional regression estimators, such as functional local linear estimator of \cite{BFR+07} or $k$-nearest neighbour estimator of \cite{BFV09}. The functional local linear estimator can improve the estimation accuracy of the regression function by using a higher-order kernel function. The $k$-nearest neighbour estimator takes into account the local structure of the data and gives better forecasts when the functional data are heterogeneously concentrated.
  \item Extend to nonparametric functional regression model that admits the mixed types of regressors with heterogeneous errors. The covariate-dependent variance can be modelled by another kernel density estimator.
  \item Extend to nonparametric functional regression model that admits the mixed types of regressors with autoregressive errors.
\item Extend to semi-functional partial linear model that admits the mixed types of regressors with homogenous, heterogeneous and autoregressive errors.
\item Apply the idea of marginal likelihood to select optimal semi-metric in nonparametric functional regression models.
\end{compactenum}

\newpage
\bibliographystyle{apalike}
\bibliography{master}

\end{document}